\documentclass[aps,preprint,amsmath,amssymb]{revtex4}
\usepackage{graphicx}
\begin{document}
\begin{center}
{\large\bf Quantum Criticality in Heavy Fermion Metals
}
\\[0.5cm]


Philipp Gegenwart$^{1}$, Qimiao Si$^{2}$
\& Frank Steglich$^{3}$ \\

{\em $^1$I. Physik. Institut, Georg-August Universit\"{a}t
G\"ottingen, D-37077 G\"ottingen}, Germany\\

{\em $^2$Department of Physics and Astronomy, Rice University,
Houston, TX 77005, USA}\\

{\em $^3$
Max-Planck-Institut f\"ur Chemische Physik fester Stoffe,
D-01187 Dresden, Germany}\\


\end{center}

\vspace{0.5cm}
{\bf
Quantum criticality describes the collective fluctuations of matter
undergoing a second-order phase transition at
zero temperature.
Heavy fermion metals have in recent years emerged as prototypical
systems to study
quantum
critical points.
There have been
considerable efforts, both experimental and theoretical,
which use
these magnetic systems to address
problems
that are central to the broad
understanding of strongly correlated quantum matter. Here, we summarize
some of the basic issues, including i) the extent to which
the quantum criticality in heavy fermion metals goes beyond the
standard theory
of
order-parameter fluctuations,
ii) the nature of the Kondo effect in the quantum critical regime,
iii) the non-Fermi liquid phenomena that accompany quantum criticality,
and iv) the interplay between quantum criticality and
unconventional superconductivity.
}



In the classical world, matter in equilibrium freezes at absolute
zero temperature. A macroscopic collection of microscopic particles
adopt
a stationary arrangement,
forming
an ordered pattern,
to minimize
the
potential energy. Quantum mechanics, on the other hand,
allows fluctuations even at zero temperature. The effective strength
of
such
zero-point motion can be tuned through the variation
of a non-thermal control parameter, such as applied pressure.
When such quantum fluctuations become
sufficiently strong, the system
undergoes a quantum phase transition to
a new ground state.

As simple as this sounds, quantum phase transitions are not easy
to achieve. Consider, for example, the
case of ice.
Anybody who has skated
would appreciate the fact that
the melting temperature of ice is reduced by pressure.
If
the melting temperature
were forced to vanish at a sufficiently high pressure, a quantum
phase
transition would take place at this pressure. However,
applying pressure larger than about 0.2 GPa to ice
actually makes the melting temperature
go up again.\cite{ice}

\begin{figure}[!ht]
\begin{center}
\vskip 0.5 cm
\includegraphics[angle=0,width=0.75\textwidth]{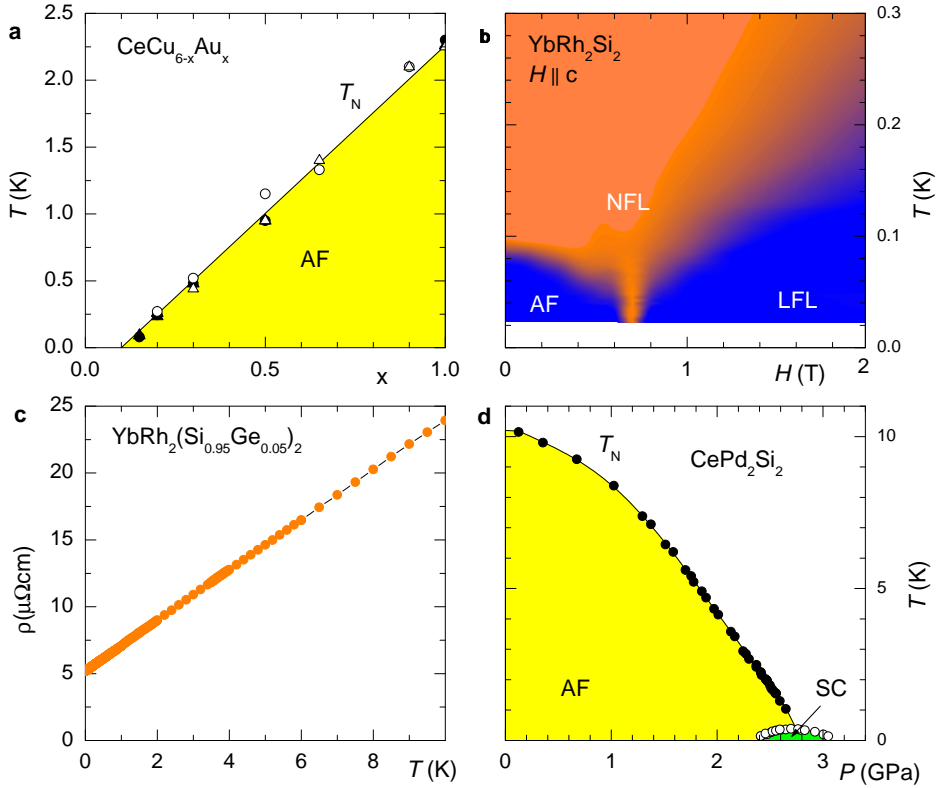}
\end{center}
\caption{Quantum critical points in heavy fermion metals.
a: AF ordering
temperature $T_N$ vs. Au concentration $x$ for CeCu$_{6-x}$Au$_x$
(Ref. \cite{Schroeder}), showing a doping induced QCP. b:
Suppression of the magnetic ordering in YbRh$_2$Si$_2$ by a magnetic
field. Also shown is the evolution of the exponent $\alpha$ in
$\Delta\rho \equiv [\rho(T)-\rho_0 ]\propto T^\alpha$, within the
temperature-field phase diagram of YbRh$_2$Si$_2$ (Ref.
\cite{Custers}). Blue and orange regions mark $\alpha=2$ and $1$,
respectively. c: Linear temperature dependence of the electrical
resistivity for Ge-doped YbRh$_2$Si$_2$ over three decades of
temperature (Ref. \cite{Custers}), demonstrating the robustness of
the non-Fermi liquid behavior in the quantum critical regime. d:
Temperature vs. pressure phase diagram for CePd$_2$Si$_2$,
illustrating the emergence of a superconducting phase centered
around the QCP.
The N\'{e}el- ($T_N$) and superconducting ordering temperatures
($T_c$) are indicated by closed and open symbols, respectively.
\cite{Mathur}}\label{overall}
\end{figure}

The transformation between ice and water also
illustrates another concept in phase
transitions. At the melting point, ice abruptly turns into water,
absorbing latent heat. In other words, the transition is of first
order. A piece of magnet, on the other hand, typically ``melts''
into a paramagnet through a continuous transition: The magnetization
vanishes smoothly,
and no latent heat is involved.
In the case of zero temperature, 
the point of such a second-order phase transition
is the quantum
critical point (QCP).\cite{Sachdev_book,Coleman-Schofield}
The quantum critical state
is distinct from the phases on both sides,
and is expected to display features in its physical properties
that are universal. Moreover, 
it will contain emergent low-energy excitations that are 
highly collective, thereby representing
a quantum state of matter with properties that are necessarily
different from those of any weakly interacting system.

It is not until recent years that QCPs have been experimentally
observed. \cite{Stewart_01,Loehneysen_07}
Particularly clear-cut examples have
come from heavy fermion (HF) metals, rare-earth-based intermetallic
compounds
in which the
effective charge carrier masses
are hundreds times that of the bare electron mass.
Accompanying the
large effective electron mass is the fact that the relevant energy
scales in the
HF metals are so small, that their ground states can be readily
tuned not only by chemical doping but also through a range of
pressure or magnetic field accessible in many condensed-matter
physics laboratories.

Two examples best illustrate the observation of QCPs in HF metals.
In CeCu$_{6-x}$Au$_x$ (Refs.\cite{Loehneysen,Schroeder}), Au-doping
turns the paramagnetic metal phase of the pure CeCu$_6$ into an
antiferromagnetic (AF) metal phase for
$x > 0.1$
(Fig.~1a).
As a function of the doping concentration,
the AF transition temperature
($T_N$)
sets in continuously;
the same is the case for the
low-temperature staggered moment -- the spatially-modulated magnetization
at the AF wave vector.
These features
establish the existence
of an AF QCP.
In YbRh$_2$Si$_2$, nature is even kinder because the
magnetic transition temperature
is as low as 70~mK in the parent undoped compound. \cite{Trovarelli}
The magnetic ordering is continuously
suppressed
by the
application of a small magnetic field, leading to a field-driven AF
QCP.
Of course, measurements can never be extended down to zero temperature.
Our statement about the continuous nature of the transition
is made based on the
understanding that the lowest measured temperature, 50~mK and 20~mK
in CeCu$_{6-x}$Au$_x$ ($x \ge 0.1$)
and YbRh$_2$Si$_2$ respectively, is
sufficiently low.
(This is to be compared with the upper temperature limit of the
quantum-critical scaling regime, which is about 5~K and 10~K
in the two systems, respectively.)
Similar understanding is assumed when related statements are
made henceforth.

The explicit identification of the QCPs in these and related HF
metals has in turn helped to establish a number of properties that
are broadly important for the physics of strongly correlated
electron systems.
One of the modern themes, central to a variety of strongly correlated
electron systems, is how the standard theory of metals,
Landau's Fermi liquid theory, can break down (see below, Sec.~I).
Quantum criticality, through its emergent excitations,
serves as a mechanism for non-Fermi liquid (NFL) behavior,
as demonstrated by a $T$-linear electrical resistivity (Figs.~1b
and 1c). Moreover, the
NFL
behavior
covers a surprisingly large part of the phase diagram.
For instance, in Ge-doped YbRh$_2$Si$_2$, the $T$-linear electrical
resistivity extends over three decades of temperature (Fig.~1c),
a range which contains a large entropy (see below). Finally,
quantum criticality
can lead to
novel quantum
phases such as unconventional superconductivity (Fig.~1d).

These experiments have mostly taken place over the past decade,
and
they have been accompanied by extensive theoretical studies. The
latter have led to two classes of quantum criticality for heavy
fermion metals. One type extends the standard theory of second-order
phase transitions to the quantum case,\cite{Hertz,Millis,Moriya}
whereas the other type invokes
new
critical excitations that
are inherently quantum-mechanical.\cite{Si,Coleman,Senthil} The
purpose of this article is to provide a status report on this
rapidly developing subject.

\section{Magnetic heavy fermion metals and
Fermi liquid behavior}
\label{heavy_fermion_intro}

HF phenomena
first appeared
in the low-temperature
thermodynamic and transport properties of CeAl$_3$
in 1975.\cite{Andres}
The 1979 discovery of superconductivity in
CeCu$_2$Si$_2$ \cite{Steglich 79}
made HF physics a subject of extensive studies.
This discovery was initially received by the community with strong
scepticism which, however,
was
gradually
overcome with
the aid of
two observations, {\it i.e.} of (i)
bulk
superconductivity in high-quality CeCu$_2$Si$_2$ single crystals,
\cite{Assmus} and (ii)
HF superconductivity in
several
U-based
intermetallics: UBe$_{13}$, \cite{Ott 83} UPt$_3$, \cite{Stewart
UPt3} and URu$_2$Si$_2$. \cite{Schlabitz}
Around the same time, it was recognized that
CeCu$_2$Si$_2$, CeAl$_3$ and other Ce-based compounds behaved as
``Kondo-lattice'' systems.\cite{Grewe}

\subsection{Kondo effect}

Consider a localized magnetic moment of spin
$\frac{\hbar}{2}$ immersed in
a band of conduction electrons.
The Kondo interaction -- an exchange coupling between
the local moment and the spins of the conduction
electrons -- is AF.
It
is energetically favorable for the
two types of spins
to form an up-down
arrangement: when the local moment is in its up state, $|\uparrow>$,
a linear superposition of the conduction-electron orbitals will be in
its down state, $|\downarrow>_c$; and vice versa. The correct ground
state is not either of the product states, but an entangled state --
the Kondo singlet, $\frac{1}{2}$($|\uparrow>|\downarrow>_c -
|\downarrow>|\uparrow>_c$). One of the remarkable features is that
there is a Kondo resonance in the low-lying many-body excitation
spectrum. The singlet formation in the ground state turns a
composite object, formed out of the local moment and a conduction
electron, into an elementary excitation with internal quantum
numbers that are identical to those of a bare electron -- spin
$\frac{\hbar}{2}$ and charge $e$.
Loosely speaking,
because of the entanglement of the
local moment with the spin degree of freedom of a conduction electron,
the local moment has acquired all the quantum numbers of
the latter and is
transformed
into a composite fermion.
We will use the term Kondo effect to describe the phenomenon
of Kondo-resonance formation at low temperatures.

At high temperatures, on the other hand, the system wants to
maximize the entropy by sampling all of the possible configurations.
It gains free energy by making the local moment to be essentially
free, which in turn weakly scatters the conduction electrons; this
is the regime of asymptotic freedom, a notion that also plays a
vital role in quantum chromodynamics.
It is in this regime that
Kondo discovered
logarithmically-divergent correction terms in the scattering
amplitude beyond the Born approximation. \cite{Kondo} Kondo's work
opened a floodgate to a large body of theoretical work \cite{Hewson}
which, among other things, led to a complete understanding of the
crossover between the high temperature weak-scattering regime to the
low-temperature Kondo-singlet state. This crossover occurs over a
broad temperature range, and is specified by a Kondo temperature;
the latter depends on the Kondo interaction and the density of
states of the conduction electrons at the Fermi energy. We will use
Kondo screening to refer to the process of developing the Kondo
singlet correlations as temperature is lowered.

\subsection{Kondo lattice and heavy Fermi liquid}

HF metals contain a lattice of strongly correlated $f$-electrons and
some bands of conduction electrons. The $f$-electrons are associated
with the rare-earth or actinide ions
and are, by themselves,
in a Mott-insulating
state:
The on-site Coulomb repulsion is so
much stronger than the kinetic energy
that these $f$-electrons
behave as localized magnetic moments, typically at room temperature
and below. They are coupled to the conduction electrons via
an (AF) Kondo interaction.
In theoretical model studies, only one band of conduction electrons
is typically considered.
Such a coupled system is
called a Kondo lattice.

It is useful to
compare the HF metals
with other
strongly correlated electron systems. The Mott-insulating
nature of the $f$-electrons not only provides the connection
to the single-impurity Kondo problem, but also
highlights the similarity between
the
HF metals and doped Mott-insulators occurring in transition
metal oxides -- such as high temperature superconductors.
The main difference between
the HF metals and the doped Mott insulators lies in the way the
itinerant charge carriers are introduced to cause delocalization of
the $f$- or $d$-electrons.
In HF metals,
delocalization of the Mott-insulating $f$-electrons arises due to
their coupling to separate bands of conduction electrons. In
transition metal oxides, it is common to think of 
such a delocalization of $d$-electrons
in terms of
the addition/removal of
electrons onto/from the
transition-metal sites of the Mott insulator.

Historically, the HF metals started out as a fertile ground to study
strongly correlated Fermi liquids and
superconductors. \cite{Grewe}
FL theory
describes an interacting electron system in terms
of Landau quasiparticles, that is, dressed electrons which are
adiabatically
connected to bare electrons when the electron-electron interactions are
adiabatically turned off. These quasiparticles
are
long-lived at low energies, and have the same
internal quantum numbers as
bare electrons.

In a Kondo lattice, as in a single-impurity Kondo problem,
the entanglement in the Kondo-singlet ground state
leads to
Kondo resonances, which
constitute the Landau quasiparticles of
the resulting FL state.\cite{Hewson}
In this way, the local moments should be counted in
the Fermi volume, and the corresponding Fermi surface
is ``large''. \cite{Hewson}
The quasiparticles have a heavy mass: After all, they come from
magnetic moments which were spatially localized in the absence of
the Kondo effect. They also experience some residual interactions,
described by the Landau parameters.
From a theoretical point of view, the strongly
correlated FL state of a Kondo lattice has served as a testing
ground for many-body methods -- a very successful
recent one being the dynamical
mean field theory \cite{KotliarVollhardt}.

The most common characteristics of a FL state in a clean metal include
a $T^2$ term as the leading temperature dependence of the electrical
resistivity $\Delta \rho$($=\rho (T) -\rho_0$), and a $T$-linear term
as the leading temperature dependence of the electronic specific heat.
By enhancing the amplitude of these terms, the heavy carrier mass
makes it much easier to observe these FL properties in HF metals than
in simple metals and transition-metal compounds.

\subsection{Towards quantum criticality in heavy fermion metals}

In addition to the Kondo interaction, there also exists the
Ruderman-Kittel-Kasuya-Yoshida (RKKY) interaction between the local
moments. At the level of energetics, the interplay between these two
kinds of interactions was
emphasized early on.\cite{Doniach,Varma76} Doniach, in particular,
considered a Kondo-lattice model that is simplified in several ways:
It lives in one dimension (1D) and deals with a purely insulating case
-- the conduction electron band is
replaced by a lattice of coupled localized-spins.\cite{Doniach} Based on
energetics, he concluded that this model has a second-order quantum
phase transition: The ground state is an insulating paramagnet when
the Kondo interaction dominates over the RKKY interaction, and an
insulating antiferromagnet when the RKKY interaction is larger
instead. Years later, we have understood that some of the quantum
fluctuation effects,
which are especially strong in 1D
and neglected in
the initial
treatment, would
in fact lead to a very different
phase diagram for the
particular 1D model.
Still, the competition between the two types of
interactions in any Kondo-lattice model is by now well established
to be an important microscopic ingredient
of HF systems.

Some of the earliest clues pointing to the potential failure of
the FL description came from
certain
alloyed
HF materials. \cite{Seaman 91,Andraka 91}
However, it
is
the study of the
(nearly) stoichiometric HF metals which have provided the direct
linkage between NFL phenomenology and quantum criticality.

It is perhaps appropriate to consider the study of quantum
criticality, following that of superconductivity, as the second
revolution of the HF field. HF metals have been used as a model
setting to study QCPs and have elucidated the richness of quantum
criticality. At the same time, the notion of quantum criticality is
forcing us to revisit some of the age-old questions about HF physics
{\it per se}, including the emergence of heavy quasiparticles and
the interplay between magnetism and superconductivity.




\section{\bf Nature of quantum criticality}

How do we theoretically
describe quantum criticality?
In Landau's theory of phase transitions,
phases
are distinguished in terms of
an order parameter, which characterizes symmetry breaking.
For instance, a staggered magnetization differentiates
between an antiferromagnet and a paramagnet.
At a second-order phase transition,
the
critical fluctuations
are characterized by the spatial fluctuations of the order
parameter.
The length scale over which such
fluctuations are correlated defines the correlation length, $\xi$,
which diverges at the critical point.
The order-parameter fluctuations
are described in terms of a field theory, commonly called a $\phi^4$-theory,
in $d$-spatial dimensions.

\begin{figure}[!ht]
\begin{center}
\vskip 0.5 cm
\includegraphics[angle=0,width=1.00\textwidth]{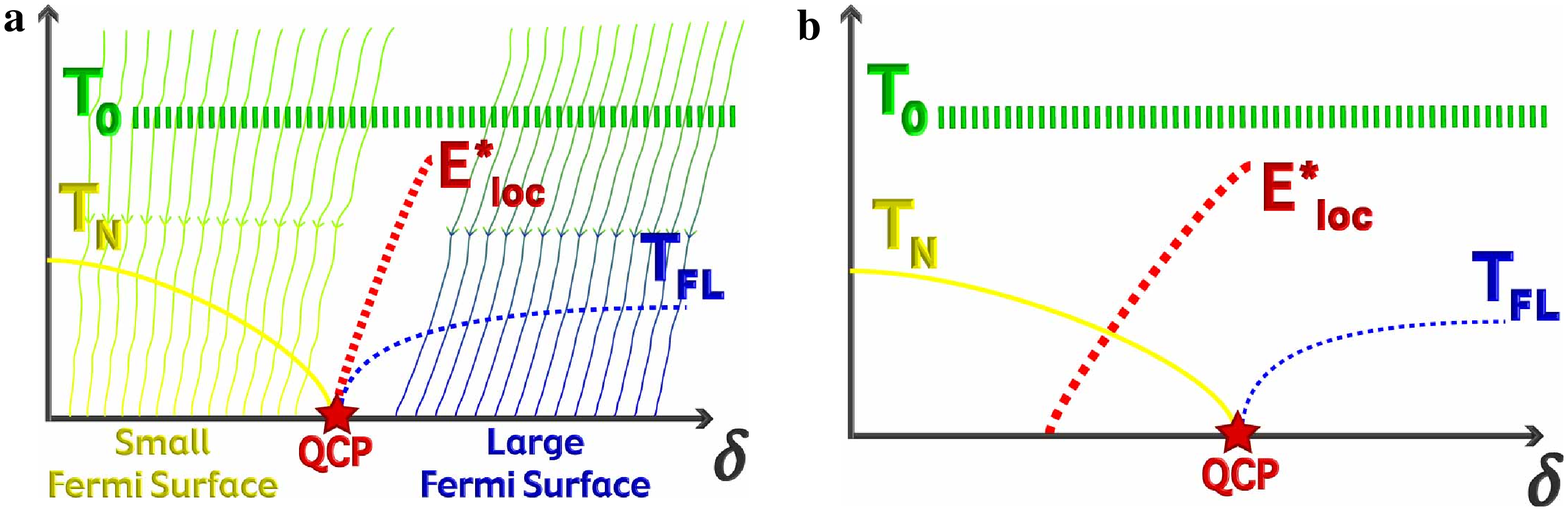}
\end{center}
\caption{Schematic phase diagrams displaying two classes of quantum
critical points.
Shown are the temperature/energy
scales vs. control-parameter ($\delta$,
which tunes
the ratio of the
Kondo
interaction
to
 the RKKY interaction),
illustrating quantum
criticality with critical Kondo destruction (a) and of the spin-density-wave
type (b).
$T_N$ represents the N\'{e}el temperature and
$T_{\text{\small FL}}$
the onset of the low-temperature Fermi liquid regime.
Lines with arrow correspond to renormalization-group flows,
which describe the transformation of the system from the high
temperature fully-incoherent regime to the zero-temperature
ground states.
$E_{\rm loc}^*$
marks
an energy scale separating the renormalization-group flows
towards two types of ground states -- one with a large Fermi
surface
(Kondo
resonance fully
developed,
and $f$-electrons delocalized)
and the other with a ``small'' Fermi
surface
(static Kondo screening absent,
and $f$-electrons localized).
Similar renormalization-group flows apply to (b), but are
omitted there
for visual simplicity.
$T_0$ signifies the initial crossover in a Kondo lattice system,
from the high temperature regime, where the local moments are completely
incoherent, to the intermediate temperature regime in which the
initial Kondo screening operates.
}
\label{theory}
\end{figure}

The standard description of a
{\it quantum} critical point
follows this prescription. The
critical mode is still considered to be the fluctuations of the
order parameter, a classical variable. The only vestige of the
quantum nature of the critical fluctuations is that they take place
not only in space, but also in imaginary time. \cite{Hertz,Chakravarty_89}
The
correlation time, $\tau_0$, scales with the correlation length,
$\xi$, defining a dynamic exponent $z$, via $\tau_0 \propto \xi^z$.
For an AF metal, the result is a $\phi^4$-field theory, in
$d+z$-spatial dimensions, with $z=2$. The finite temperature
properties of this theory have been systematically discussed.
\cite{Millis,Moriya,Continentino} More microscopically, an itinerant
antiferromagnet
is typically described in terms of a spin-density-wave state, {\it
i.e.} a spontaneous
spatial
modulation of the spins of the charge carriers,
and the picture of quantum critical order-parameter fluctuations in
an itinerant antiferromagnet
is also called a spin-density-wave QCP.

Theoretical studies in recent years
have led to the
notion
that the
order-parameter-fluctuation
description may be insufficient,
at least in some cases.
Instead, inherent quantum effects
play an
important role.\cite{Si,Coleman,Senthil,Senthil_deconfined}
In these cases,
one needs to identify the additional critical modes before a
critical field theory can be constructed. The systems being
considered in this context
include
not only
HF
metals, but also quantum
insulating magnets.
There is so far
no unified framework to identify the new critical modes,
and the different systems are
being studied theoretically
on a case by case basis.

In
HF metals, the additional critical modes have been described in
terms of the destruction of local Kondo singlets at the $T=0$ AF
transition. In the
framework
of ``local'' quantum criticality, \cite{Si,Si_03} the
Kondo effect is critically
destroyed because local moments are coupled not only
to the conduction electrons but also to the
fluctuations
of the other local moments.
Such magnetic fluctuations,
whose spectrum turns critically soft
at the QCP,
act as a source of dissipation
and decohere the Kondo effect.
This critical destruction of the Kondo effect
implies that the magnetic transition is accompanied by
a continuous localization-delocalization
transition of the electronic excitations.
A separate mechanism has been postulated 
in the form of spin-charge
separation. \cite{Coleman} Finally, spin-liquid formation
among the local moments,
which may
occur as a result of geometric frustration, has been proposed as a
competing mechanism against Kondo-singlet formation. \cite{Senthil,Georges}

It is worth emphasizing the dynamical nature of the Kondo screening,
{\it cf.} Fig.~\ref{theory}a.
At sufficiently high temperatures, a Kondo lattice
behaves as a collection of individual local moments, weakly coupled
to a band of conduction electrons. At the temperature $T_0$,
the initial crossover into Kondo screening sets in.
We should caution that extracting $T_0$ from the 
Kondo temperature of the diluted limit of 
a heavy fermion system is in general problematic,
since the process of dilution can change the atomic volume \cite{FS}
and other parameters. Still, the initial crossover associated with
$T_0$ of the Kondo lattice system has a similar physical meaning as
that of the crossover occurring in a single-impurity Kondo problem.
For definiteness, and inspired by the exact solution to the
thermodynamics of the single-impurity Kondo problem, \cite{Melnikov}
we define $T_0$ as twice of the temperature at which the entropy is
0.4~Rln2 per local moment.

Whether the Kondo screening process is complete or incomplete {\it
at sufficiently low temperatures} depends on the competition between
the Kondo interaction and the RKKY coupling among the moments. The
line associated with the energy scale $E_{\text{\small loc}}^*$
is the separation between the two
regimes.
Upon increasing $\delta$ at a fixed temperature
(Fig.~\ref{theory}a),
the $E_{\text{\small loc}}^*$ line
marks the crossover from the small
Fermi-surface regime to the large-Fermi-surface one.
To the right of the $E_{\text{\small loc}}^*$ line
the local moments are being converted into the composite
fermions, which are part of the Fermi volume. In this regime one finds,
upon cooling at a fixed $\delta$, that FL coherence is fully established
below $T_{\text{\small FL}}$.
In the regime to the left of the $E_{\text{\small loc}}^*$ line,
the Kondo screening is incomplete even at zero temperature, and
the local moments do not participate in the Fermi-surface formation.
Here, the system eventually orders into an AF state below the N\'{e}el
temperature, $T_N$; at the lowest temperatures inside the AF phase,
the system will also be a Fermi liquid (see below, Sec.~\ref{phases},
for further discussions).
In the zero-temperature limit,
the $E_{\text{\small loc}}^*$ line marks a genuine
f-electron-localization phase transition; at
finite temperatures, it represents a crossover.

When $E_{\text{\small loc}}^*$ terminates at the same value
of the
control parameter ($\delta_c$) as the AF phase boundary
(Fig.~\ref{theory}a), the quantum
critical fluctuations include not only the fluctuations of the
magnetic order parameter but also those associated with the
destruction of Kondo
effect.
The local quantum critical solution belongs to
this case.\cite{Si,Si_03,Zhu_03,Zhu07,Glossop07}
When the two lines intersect
(Fig.~\ref{theory}b), the
magnetic QCP falls in the category of the spin-density-wave theory.
The third case, with the $E_{\text{\small loc}}^*$ line
terminating before meeting the magnetic-ordering phase boundary,
would imply a finite range of $T=0$ parameter space in which
the local moments are neither Kondo-screened nor magnetically
ordered; this does not happen in general, although it cannot be
completely ruled out if the underlying spin system is highly frustrated.


\section{\bf Quantum critical scaling}

Historically, scaling has played a central role
in the study of classical
criticality. It results from the non-analyticity in the free energy
at
the transition temperature $T_c$, and is manifested
in thermodynamic properties
such
as specific heat and static susceptibility.

Quantum criticality
shows a number of
important distinctions in scaling.
First, there are generically two control parameters which are both
relevant in the renormalization-group sense.
One is the non-thermal control parameter, $\delta-\delta_c$, and the
other one is the temperature $T$. They represent two independent
directions to access the QCP ($T=0, \delta = \delta_c$) within the
temperature-control-parameter phase diagram. Accessing the QCP by
varying the non-thermal control parameter will lead to a divergence
of the spatial correlation length, $\xi$. Doing so through reducing
the temperature $T$, on the other hand, amounts to increasing the
size of the temporal dimension, $\beta = 1/k_BT$, towards infinity.
Even the scaling
of thermodynamic quantities would involve judicious
mixing of spatial and temporal fluctuations. This is in contrast to
the case of a generic classical critical point, where the
non-thermal control parameter and the reduced temperature
$(T-T_c)/T_c$ are inter-dependent; the variation of each will be
manifested in scaling through the spatial correlation length.
Second, quantum critical
scaling is also delicate due to the very fact that a QCP
exists
at zero temperature. The system is therefore subject to the
constraints of the third law of thermodynamics. For instance, a
divergence of the specific heat often occurs at a classical critical
point, but cannot occur at a QCP.

It turns out that
static-scaling exponents are more conveniently defined in terms of
thermodynamic ratios. A prototypical example is the Gr\"uneisen
ratio -- {\it defined} \cite{zhu} as the ratio of the volume thermal
expansion
to the specific heat,
$\Gamma=\beta/C_p$. At a generic QCP where the control
parameter is linearly coupled to the pressure, $\Gamma$ must
diverge, in a way that provides a thermodynamic means of probing the
universality classes. \cite{zhu}
It follows from straightforward scaling arguments that
the critical Gr\"uneisen ratio, {\it i.e.} the ratio of the
critical part of thermal expansion
to
the critical part of specific heat, goes as $\Gamma^{cr} \sim
T^{-x}$ at the QCP.
The Gr\"uneisen exponent $x= 1/\nu z$, with $\nu$ being
the correlation-length exponent [via $\xi \propto (\delta - \delta
_c)^{-\nu}$] and $z$ the dynamic exponent. For a magnetic-field
driven transition, similar
result
applies to the magnetocaloric
effect $(1/T)\partial T/\partial H |_S$. \cite{zhu,garst}

\begin{figure}[!ht]
\begin{center}
\vskip 0.5 cm
\includegraphics[angle=0,width=0.75\textwidth]{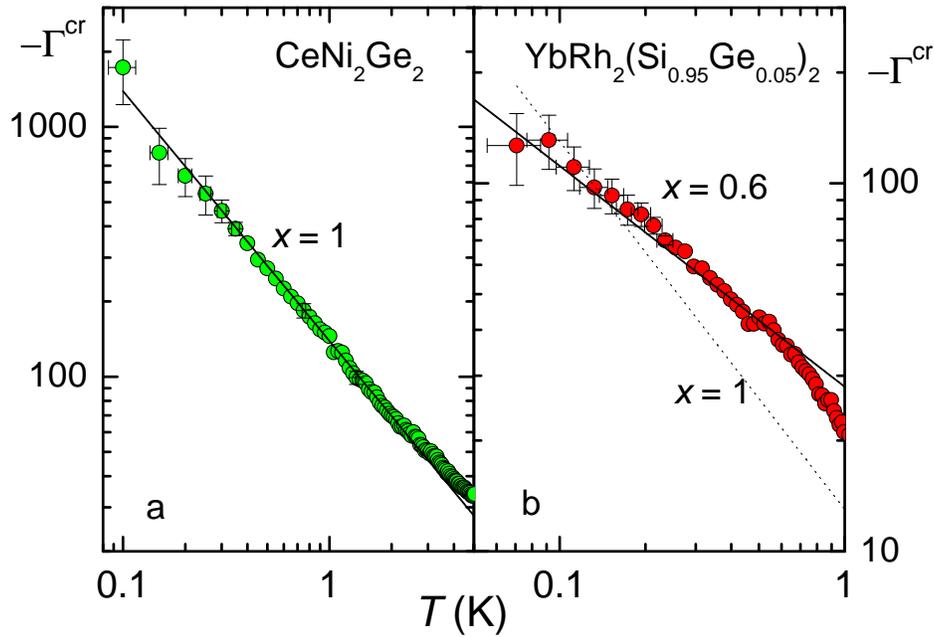}
\end{center}
\caption{Divergence of the dimensionless critical Gr\"{u}neisen
ratio at quantum critical points.
Plotted are the temperature dependence of
$\Gamma^{cr}=V_m/\kappa_T\times \beta^{cr}/C^{cr}$
for CeNi$_2$Ge$_2$ (a) and YbRh$_2$(Si$_{0.95}$Ge$_{0.05}$)$_2$ (b),
on double-log scales.\cite{Kuechler 2003}
Here, $\beta^{cr}$
and $C^{cr}$ are the volume thermal expansion and specific heat
after subtraction of normal (Fermi-liquid) contributions.
In addition, $V_m$ and $\kappa_T$ are the molar volume and
isothermal compressibility (at room temperature); they are
used as normalization to make $\Gamma^{cr}$ dimensionless.
Error
bars, standard errors.}\label{fig3}
\end{figure}

Divergence of the critical Gr\"uneisen ratio has now been observed in several
HF metals near their QCP. Fig.~\ref{fig3} shows the results for
single crystals of CeNi$_2$Ge$_2$ and
YbRh$_2$(Si$_{0.95}$Ge$_{0.05}$)$_2$. For CeNi$_2$Ge$_2$, the
Gr\"uneisen exponent $x$ is consistent with a spin-density-wave QCP
in three dimensions (3D), which predicts $x=1$.\cite{Kuechler 2003}
Similar observations
have also been made for some other systems. \cite{Kuechler
2006,Gegenwart Sr327} By contrast, the exponent $x \approx 0.6$ in
YbRh$_2$(Si$_{0.95}$Ge$_{0.05}$)$_2$ (Ref. \cite{Kuechler 2003})
cannot be explained by any spin-density-wave-QCP
theory.
(The exponent is roughly compatible with the
ferromagnetic paramagnon
picture which, however, is not compatible with other experimental
results. \cite{Gegenwart 2007})
It is instead compatible with the local
quantum critical picture
in the presence of xy-spin anisotropy, which has $x \approx 0.66$
[Zhu, L. \& Si, Q. unpublished (2003); Zhu, L.
Ph. D. Thesis, Rice University (2005)].
A phenomenological approach\cite{Pepin05}
has also yielded a fractional Gr\"uneisen exponent $x=2/3$, when
an emergent spinless-fermionic field is assumed to take a
certain fine-tuned dispersion; this picture, though, contains no
Fermi-surface jump at the QCP. The Gr\"uneisen exponent for
CeCu$_{6-x}$Ag$_x$ also deviates from the spin-density-wave
prediction. \cite{Kuechler 2004}

The mixing of statics and dynamics inherent to any QCP implies that
the scaling pertinent to the universality class can also be observed in
dynamical measurements.
A spin-density-wave QCP in 3D is described by a five-dimensional
$\phi^4$ field theory, which has a Gaussian fixed point. The
non-interacting nature of the fixed point implies that spin damping
is absent at the fixed point, and that the corresponding relaxation
rate must be proportional to
$T^y$ with $y>1$; this means, there is a violation of
$\omega/T$ scaling in the dynamical spin
susceptibility.\cite{Sachdev_book,Moriya,Millis} For
any
interacting fixed point, on the other hand, spin damping occurs
already at the fixed-point level, and an $\omega/T$ scaling is
expected.\cite{Sachdev_book,Varma_MFL,Aronson}

In the case of CeCu$_{5.9}$Au$_{0.1}$,
extensive
measurements
with
inelastic neutron scattering have been carried out.
Most surprisingly, critical scattering occurs along certain lines
(as opposed to points) in ${\bf q}$-space,
indicating quasi-two-dimensional (2D) quantum
critical fluctuations. \cite{Stockert}
However, the 2D spin-density-wave QCP
picture cannot
explain
the observation,
that both at and far away from the wave vector of the nearby
AF order, the neutron and magnetization data exhibit $\omega/T$
scaling,
with an anomalous exponent $\alpha<1$ (Ref.
\cite{Schroeder}).
(The same exponent appears in the $H/T$ scaling of
the magnetization of CeCu$_{5.9}$Au$_{0.1}$.)

Considerable theoretical efforts
have
gone into
understanding this scaling behavior.
Within the local quantum critical
picture,
the emergent critical excitations associated with
the Kondo destruction make the fixed point to be an
interacting one,
and
the
dynamical susceptibility
displays an $\omega/T$-scaling.\cite{Si,Si_03}
In
the Ising-anisotropic case, appropriate for the Au-doped
CeCu$_{6}$ system, the fractional exponent $\alpha$ takes
a value \cite{Zhu_03,Zhu07,Glossop07}
close to the experimental
observation.

In YbRh$_2$(Si$_{1-x}$Ge$_x$)$_2$ ($x=0$ and $x=0.05$),
neutron-scattering results on the critical fluctuations (as well as
on the magnetic order) are not yet available. Still, a fractional
exponent has been seen in the bulk linear-response susceptibility.
From 10~K to 0.3~K, $\Delta\chi$ is proportional to $T^{-0.6}$,
where $\Delta\chi$ denotes the susceptibility after subtracting a
small temperature independent contribution. \cite{Geg 05} This
points to some degree of universality between YbRh$_2$Si$_2$ and
CeCu$_{5.9}$Au$_{0.1}$. At the same time, there also exist some
important differences between these two systems.
The bulk susceptibility of YbRh$_2$(Si$_{1-x}$Ge$_x$)$_2$ is very
strongly enhanced at low temperatures. Related
behavior is also found in the field dependence of the ($T\rightarrow
0$) Pauli susceptibility $\chi_0$ at $H>H_c$. In addition, the
Sommerfeld-Wilson ratio of $\chi_0$
to
the electronic specific heat
coefficient, $\gamma=C_{el}/T$, is extremely enhanced compared to
usual
HF
metals and exceeds a value of
$30$
on approaching
the critical field. \cite{Geg 05} These observations
have been
described
in terms of a very small Weiss field
at ${\bf q=0}$ (Ref. \cite{Gegenwart 2007}).
A recent proposal\cite{Imada}
associates the enhanced ferromagnetic fluctuations
with a possible nearby quantum tricritical point.
Enhanced
ferromagnetic correlations
might also
be responsible for the sharp Yb$^{3+}$-ESR line observed well
below $T_0$.\cite{Sichelschmidt}

The issue of violation of $\omega/T$ scaling
in candidate spin-density-wave QCPs
has not yet been extensively studied. As will be discussed in
Sec.~\ref{potential_qcp},
some indication along this
direction has appeared in lightly Rh-doped
CeRu$_2$Si$_2$. \cite{Kadowaki}


\section{\bf Divergence of effective mass}
\label{divergence_mass}

\begin{figure}[!ht]
\begin{center}
\vskip 0.5 cm
\includegraphics[angle=0,width=0.75\textwidth]{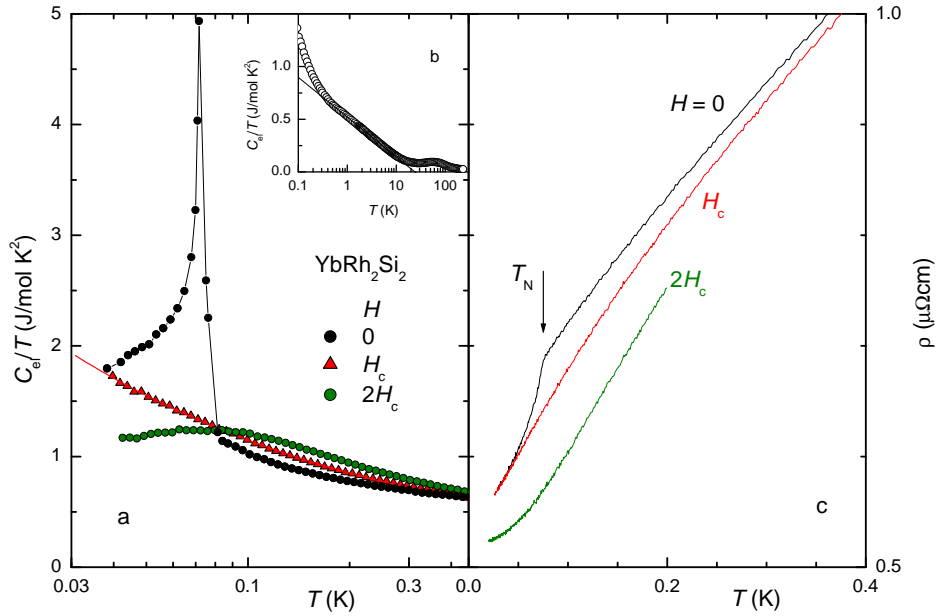}
\end{center}
\caption{Thermodynamic and transport properties close to the quantum
critical point in YbRh$_2$Si$_2$. a: Low-temperature electronic
specific-heat coefficient at various different fields applied
perpendicular to the $c$-axis [Oeschler, N. {\it et al.} {\it
Physica B}, in press (2007)]. At the critical field, $H_c \approx
0.055$~T, the specific-heat coefficient goes as $T^{-0.4}$ at low
temperatures and as $-{\rm ln}T$ at higher temperatures.
b: The zero-field data over an
extended temperature range, illustrating the onset of the ${\rm
ln}T_0/T$ dependence at $T \approx 10$~K ($T_0 = 24$~K) and the
onset of a power-law divergence at $T \approx 0.3$~K.
\cite{Trovarelli}
c: Low-temperature electrical resistivity
[Gegenwart, P. {\it et al.}
{\it Physica B}, in press (2007)]
as a function of temperature at the same magnetic fields as
in a.}\label{fig4}
\end{figure}

A key characteristic of both CeCu$_{5.9}$Au$_{0.1}$ and
YbRh$_2$Si$_2$ is the divergence of the effective charge-carrier
mass at the QCP.
The case of YbRh$_2$Si$_2$ is illustrated in Fig.~\ref{fig4}.
The magnetic field tunes the system through a QCP at $H_c$, as reflected
in the evolution of the specific heat and electrical resistivity.
Between 10~K and 0.3~K, the electronic specific-heat coefficient
$\gamma$ shows a logarithmic divergence. A similar logarithmic
divergence was also seen in CeCu$_{5.9}$Au$_{0.1}$. A feature that
has been observed in YbRh$_2$Si$_2$, but not
in
CeCu$_{5.9}$Au$_{0.1}$, is the more singular
$T^{-0.4}$ dependence found below 0.3~K.
For $H>H_c$, both the specific heat and the electrical resistivity
of 5~at\%-Ge-doped YbRh$_2$Si$_2$ display scaling over almost four
decades in $(H-H_c)/T$.\cite{Custers} Within the FL regime,
$\gamma\propto (H-H_c)^{-1/3}$. Fermi liquid theory allows us to
identify the value of $\gamma$ at zero temperature with the
effective mass of Landau quasiparticles, appropriately averaged over
the Fermi surface. Hence, the latter can be inferred to diverge as
the QCP is reached.

For CeNi$_2$Ge$_2$ \cite{Kuechler
2003} and CeCu$_2$Si$_2$, \cite{Gegenwart 1998}
by contrast,
the
effective mass
appears to be finite.
(For both compounds, low-temperature upturns are observed in the
normal-state $C_{\rm el}/T$ results. These upturns are very likely
not part of the quantum criticality. \cite{Kuechler 2003})

A divergent effective mass naturally arises
on approach of a Kondo-destroying
QCP, because a critical destruction of Kondo screening leads to the
vanishing of quasiparticle weight everywhere on the Fermi surface.
In a spin-density-wave QCP, the vanishing of quasiparticle spectral
weight occurs only near the ``hot spots'', the portions of the Fermi
surface that are connected by the AF wave vector. On all other parts
of the Fermi surface, kinematic constraints prevent the over-damped
spin fluctuations from scattering the electrons; the quasiparticles
are well defined, with a finite mass.
(This anisotropy is also manifested in the electron scattering rates,
which have the NFL and FL forms at the hot spots and cold regions,
respectively. In the clean limit, the current carried by cold electrons
short-circuits that
carried by
the hot electrons, making $\Delta \rho \propto
T^2$ even
above $T_{\text{\small FL}}$
upon cooling
near the QCP.\cite{Hlubina,Rosch})
For a 3D spin-density-wave
QCP, the Fermi-surface-averaged effective mass appearing in the
specific-heat coefficient is finite.

%

As alluded to in the introduction,
the entropy associated with the scaling regime is very large --
about $0.4R \ln 2$ per Yb moment.\cite{Trovarelli}
It is truly striking that quantum critical excitations accumulate
nearly half of the full entropy associated with the atomic physics
of the lowest 
crystal-field derived doublet of each Yb$^{3+}$-ion. This result exemplifies
how quantum criticality is able to reach up in temperature
and
influence
a large part of the
phase diagram of
strongly correlated electron
systems. \cite{Sereni 97}





\section{\bf Fermi surface and energy scales}

\subsection{\bf Collapse of Fermi surface}

\begin{figure}[!ht]
\begin{center}
\vskip 0.5 cm
\includegraphics[angle=0,width=0.75\textwidth]{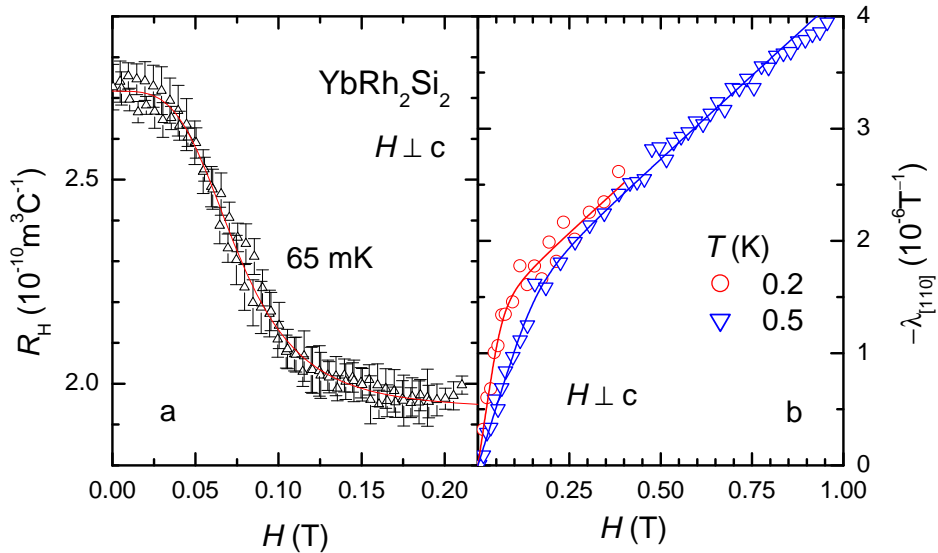}
\end{center}
\caption{Evidence for an additional low-energy scale in the
Hall effect and thermodynamic data of YbRh$_2$Si$_2$.
a: Linear-response
Hall coefficient $R_H$ as derived from the initial slope of the Hall
resistivity in a crossed field experiment
[Friedemann, S. {\it et al.} {\it Physica B}, in press (2007)],
performed on the same 
single crystal used in Ref. \cite{Paschen} Error bars, standard
errors. The crossover width decreases with temperature,
extrapolating to zero in the zero-temperature limit.\cite{Paschen}
b: Magnetic-field dependence of the magnetostriction of a
high-quality single crystal
($\rho_0 \approx 0.5~\mu\Omega$cm). The symbols
represent the linear coefficient $\lambda_{[110]}=\partial\ln
L/\partial H$ (where $L$ is the sample length along the $[110]$
direction within the tetragonal $ab$ plane) versus $H$ at different
temperatures.\cite{Gegenwart 2007} }\label{hall_lambda}
\end{figure}

The evolution of Hall effect
across
the QCP has been studied in Ref.~\cite{Paschen} for YbRh$_2$Si$_2$.
The isothermal
linear-response Hall-coefficient, as shown in
Fig.~\ref{hall_lambda}a, displays a broadened step when
the system is tuned (with
the field applied in the magnetic easy plane parallel to the current
direction) across the QCP.
This feature sharpens
upon decreasing the temperature,
extrapolating to a jump in the zero-temperature limit,
precisely at the QCP.

For a spin-density-wave QCP, the system orders into a
spin-density-wave state, which breaks translational invariance. This
order generically removes
the hot spots of the Fermi surface,
and leads to a reconstruction of the Fermi
surface. Still, the gradual onset of the spin-density-wave order
implies a smooth evolution of the Fermi surface, and the Hall
coefficient
is
continuous across the
QCP.\cite{Coleman,Norman_prl03} Precisely this type of evolution has
been seen in the Hall coefficient of V-doped Cr,\cite{Lee_prl05}
which is considered a prototypical system for a spin-density-wave
quantum phase transition.

The Hall data in YbRh$_2$Si$_2$, by contrast, suggest
a sudden change of the Fermi
surface at the QCP.\cite{Paschen} The extrapolated zero-temperature
jump, being inconsistent with the spin-density-wave QCP picture, is
naturally interpreted in terms of a collapse of the large Fermi
surface: It is as if a large number of charge carriers
of the non-magnetic phase
are suddenly
lost as we tune the system
through
the QCP. As can be seen from
Fig.~\ref{theory}, this is precisely what happens in a
Kondo-destroying QCP: The destruction of Kondo resonances at the
magnetic QCP converts the large Fermi surface of the paramagnetic HF
metal into a small one.

A
traditional
means to measure the Fermi surface is the de Haas-van
Alphen (dHvA) technique. Recent dHvA experiments \cite{Shishido}
have detected a sudden change of the Fermi surface as a function of
pressure in the field-induced normal state of CeRhIn$_{\rm 5}$ ({\it
cf.} Fig.~8). As we will discuss shortly, it is possible that the
threshold pressure seen in the dHvA measurement on CeRhIn$_{\rm 5}$
corresponds to an AF QCP. \cite{Park_06}


\begin{figure}[!ht]
\begin{center}
\vskip 0.5 cm
\includegraphics[angle=0,width=0.75\textwidth]{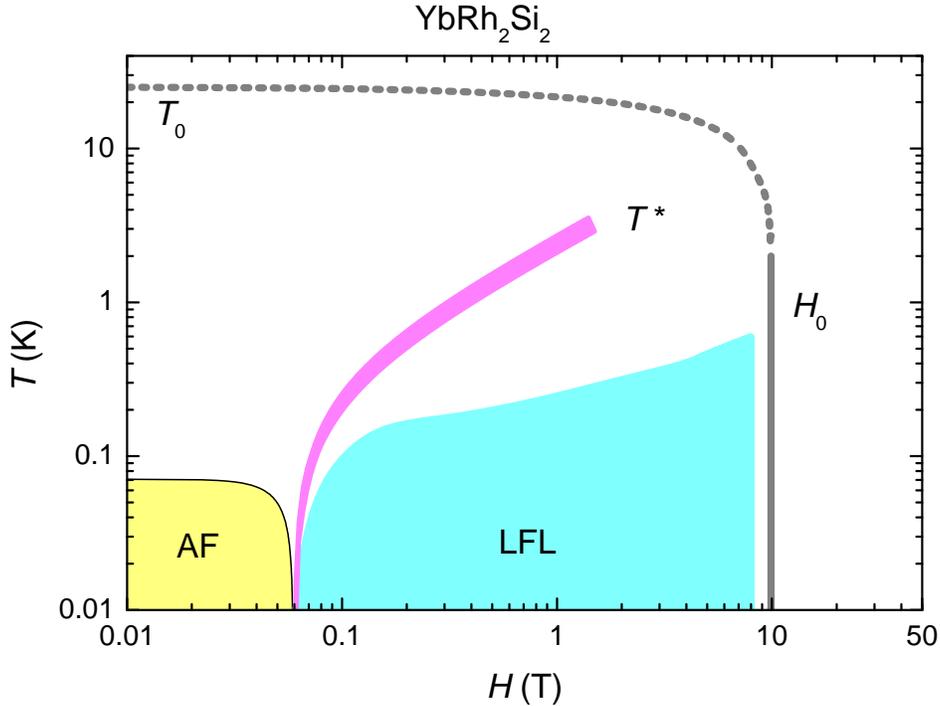}
\end{center}
\caption{The temperature vs. magnetic field phase diagram
for YbRh$_2$Si$_2$. The magnetic field ($H$) is applied along the
crystallographic
$c$-axis. Dashed and
solid gray lines display the initial Kondo crossover scale $T_0=25$~K
(at $H=0$)
and
the corresponding $H_0$,
which suppresses
the heavy quasiparticle behavior.\cite{Tokiwa 2005}
The sharp black line indicates the boundary
of the AF phase.\cite{Gegenwart 2002}
The broad red line
specifies the
position of cross-over in the isothermal Hall resistivity,\cite{Paschen}
which agrees with that
observed
in the
isothermal magnetostriction, magnetization and longitudinal
resistivity \cite{Gegenwart 2007} and marks an additional low-energy
scale. The regime
at $H>H_c$, where the electrical resistivity follows the
Landau Fermi
liquid behavior $\Delta \rho \propto T^2$, is
shown
in
blue.\cite{Gegenwart
2002}
A double-log representation has been used for clarity.
}\label{YbRh$_2$Si$_2$_overall_phase_diagram}
\end{figure}

\subsection{\bf Multiple energy scales}

The Fermi surface is sharply defined only at zero temperature.
Fig.~\ref{theory} naturally suggests that the contrast between the
two types of QCP should also be manifest in the energy scales of the
equilibrium many-body excitation spectrum and, hence, in the
finite-temperature properties as well. For the spin-density-wave
QCP, the critical modes correspond to the slow fluctuations of the
magnetic order parameter. Dynamical scaling implies that their
energy scale vanishes at the approach to the QCP. For the
Kondo-destroying QCP,  the critical Kondo effect
specifies inherently
quantum-mechanical excitations that are in addition to critical
fluctuations of the order-parameter. We can therefore expect the
occurrence of more than one energy scales, all of which vanish as the
QCP is reached. The Hall-effect measurement in YbRh$_2$Si$_2$,
described earlier, had in fact proven the existence of a
$T^\star(H)$ line
(Fig.~\ref{YbRh$_2$Si$_2$_overall_phase_diagram}).\cite{Paschen}

The Hall effect crossover is accompanied by changes in the slope of
the isothermal magnetization and magnetostriction.\cite{Gegenwart
2007} As shown in Fig.~\ref{hall_lambda}b, the latter displays a
kink-like structure whose position in field matches the position of
the Hall-effect crossover. This proves
that the low-energy scale $T^\star(H)$ is intrinsic to
the
equilibrium excitation spectrum.
The energy scale is separate from both the N\'{e}el temperature and
$T_{FL}$, the boundary of the Fermi liquid regime (cf.
Fig.~\ref{YbRh$_2$Si$_2$_overall_phase_diagram}).
The distinction between $T^*$ and $T_{FL}$ is
reinforced by the differences in how the physical properties
behave:\cite{Gegenwart 2007} various isothermal properties
display a temperature-smeared jump across the $T^*$ line,
but are always smooth across the $T_{FL}$ line.
In the $T
\rightarrow 0$ limit, the lines defined by the three energy scales
merge at the same point, the field-induced QCP.
\cite{Paschen, Gegenwart 2007}

The comparison between the thermodynamic results with
the Hall-effect data suggests that $T^\star(H)$ is the finite-$T$
manifestation of the localization of the $f$-electrons at the QCP.
%
This in turn suggests that the measured $T^*(H)$ scale
characterizes
the Fermi-surface
collapse at zero temperature.\cite{Paschen, Gegenwart 2007}
It corresponds to
the $E_{\text{\small loc}}^*$ scale of the local quantum critical
picture\cite{Si}, shown in Fig.~\ref{theory}a. Multiple vanishing
energy scales also arise in the ``deconfined'' quantum criticality
scenario for insulating quantum magnets; \cite{Senthil_deconfined}
in its extension to
itinerant
HF
systems,\cite{Senthil,Pepin}
though, the additional energy scale vanishes at a point away from
the magnetic QCP.

It is
instructive to contrast the case of
YbRh$_2$Si$_2$ with what happens in the V-doped Cr.
There, the changes in the Hall coefficient occur only at the
N\'eel transition.\cite{Yeh_nature02,Lee_prl05}
By extension, there is no indication for a separate 
$T^\star$
scale.\cite{Paschen}
These observations
are consistent with the spin-density-wave QCP description
of the V-doped Cr.

The hexagonal compound YbAgGe
is another Yb-based HF system
which undergoes field-induced
magnetic quantum phase transitions.\cite{Canfield}
It shares some similarities with
YbRh$_2$Si$_2$. In particular,
Hall-effect measurements in this material
have also shown an anomaly that defines a cross-over
line in the temperature vs. field
phase diagram.\cite{Canfield2}
Unlike for YbRh$_2$Si$_2$,
however,
the features associated with the ``Hall line''
have not yet allowed a linkage with a Fermi-surface
jump in the zero-temperature limit.
There are also some other important differences between
the two systems.
Compared to
YbRh$_2$Si$_2$,
$T_N$ in YbAgGe at ambient pressure is much
larger (about 1~K) and, correspondingly, the critical field
is much greater (about 5~T). The temperature vs. field phase
diagram is considerably more complex, with additional
first-order transitions inside the AF
region
and a
metamagnetic signature in $M(H)$ when crossing the Hall-line.
\cite{Tokiwa YbAgGe}
Thermodynamic and transport measurements
in YbAgGe have raised the possibility that
a NFL {\it phase} occurs over an extended
range of
magnetic fields.\cite{Canfield}
Given the hexagonal structure, the
underlying spin system may very well
be
frustrated; it is hence natural to raise the
question about the potential role that geometrical frustration
plays in influencing the phase diagram
of YbAgGe.

\section{\bf Superconductivity}

\begin{figure}[!ht]
\begin{center}
\vskip 0.5 cm
\includegraphics[angle=0,width=0.75\textwidth]{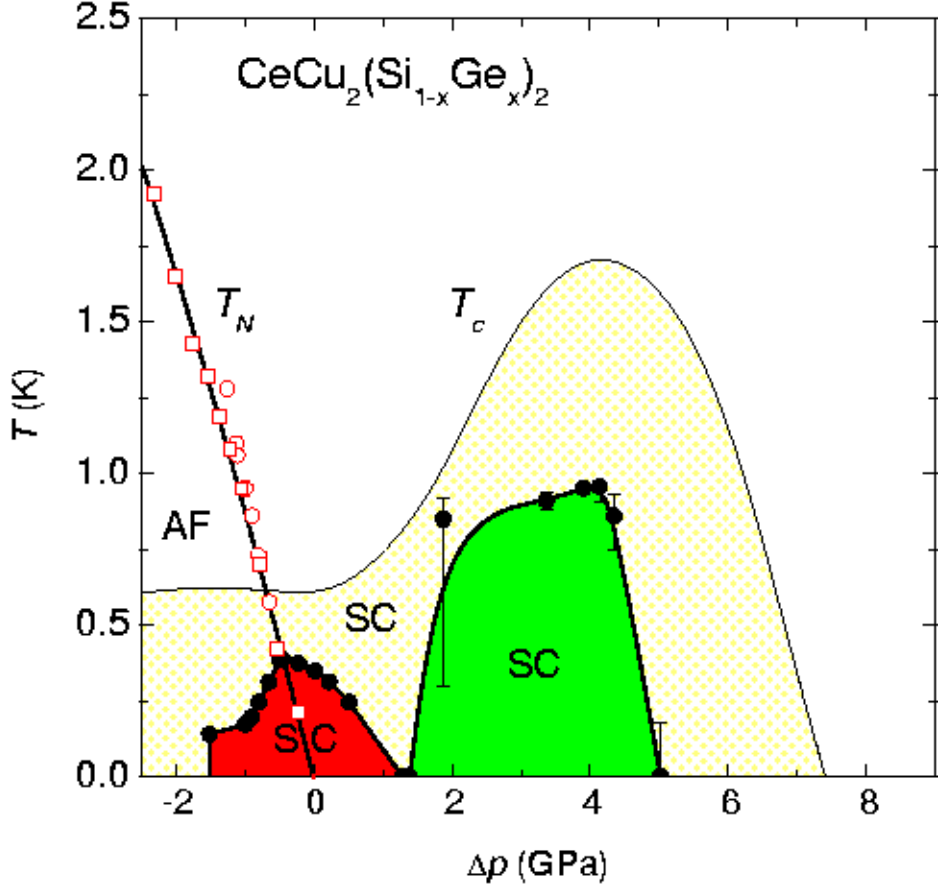}
\end{center}
\caption{Pressure dependence of magnetic order and superconductivity
in CeCu$_2$(Si$_{1-x}$Ge$_x$)$_2$.
Transition temperatures for N\'{e}el ordering $(T_N$) and
superconductivity ($T_c$) are shown as a function of pressure increment
$\Delta p=p-p_{c1}$. Error bars represent width of superconducting
transitions in the electrical resistivity. For $T_N$, the open
circles (squares) correspond to $x=0.1$ ($x=0.25$) and,
for $T_c$, the thin solid line (full circles) are for $x=0$ ($x=0.1$).
The reference pressure
$p_{c1}$ is 0.4~GPa (for $x=0$),
1.5~GPa ($x=0.1$), 2.4~GPa ($x=0.25$)
and 11.5~GPa ($x=1$).
\cite{Yuan}}\label{fig7}
\end{figure}

In the past few years growing evidence has been collected for more
than one non-phononic
pairing mechanism
operating in different
HF superconductors. These include Cooper pairing possibly mediated
by
nearly critical
valence fluctuations \cite{Miyake}
 in both CeCu$_2$Ge$_2$ \cite{Jaccard} and
CeCu$_2$Si$_2$ \cite{Yuan} under high pressure ({\it cf.} Fig.~\ref{fig7}),
by
AF magnons in UPd$_2$Al$_3$ \cite{Sato01}
and
ferromagnetic
ones
in pressurized UGe$_2$ \cite{Saxena} as well as
URhGe
\cite{Huxley} and UCoGe. \cite{Huy}

CeCu$_2$Si$_2$ (at ambient and low pressure) is
the prime candidate for
antiparamagnon-mediated
superconductivity near a
3D spin-density-wave QCP.\cite{Yuan} This is illustrated in Fig. 7
where a low-pressure dome of superconductivity occurs near the AF QCP for
10 at\% Ge-doped CeCu$_2$Si$_2$.
From
neutron-diffraction
measurements on an AF ("A-type") single crystal,\cite{Stockert_04}
the latter was found to be of the spin-density-wave variety. In
addition,
the $\Delta\rho \propto T^{1.5}$ and $\gamma(T) = \gamma_0 - \alpha
T^{0.5}$ dependences found earlier in the low-temperature normal
state in ``S-type" samples located on the strong-coupling side of the
QCP, \cite{Gegenwart 1998}
suggested that the correlated
critical fluctuations
are 3D. Ongoing inelastic neutron scattering experiments performed
on
``S-type" CeCu$_2$Si$_2$ single crystals
[Stockert, O. {\it et al.} {\it Physica B}, in press (2007)]
are
devoted to searching for quantum critical spin-density-wave
fluctuations, which were proposed to act as superconducting glue in
this case.\cite{Scalapino_86,Mathur}
Future
neutron-diffraction experiments on other NFL superconductors like
CePd$_2$Si$_2$ \cite{Mathur} are
urgently
needed to gain more insight into the relationship between
quantum criticality and superconductivity.

The
available results, however, highlight a surprisingly
rich diversity in physical scenarios.
For instance,
superconductivity
may occur from
the proximity to a
field-induced QCP,
like in CeCoIn$_5$ \cite{Petrovic} and perhaps also in
UBe$_{13}$.\cite{Gegenwart_04}

Microscopic theories of HF superconductivity were mostly
constructed before the period of active studies on quantum
criticality. Given that a large number of HF superconductors are now
recognized as being close to a QCP, it is important that the
interplay between quantum criticality and superconductivity be
systematically studied in the future. More generally, it is likely
that HF superconductors will have an important role to play in the
larger understanding of strongly correlated
superconductors.\cite{Anderson_07}

\section{\bf Discussion}

The subject of quantum critical
HF
metals is a vast one
and the present article, not meant to be comprehensive,
is unable to cover all the interesting topics.
We now touch on some of these, in the hope of pointing the
practitioners to -- and providing the uninitiated readers a glimpse
of -- some issues that are worthy of
further explorations.

\subsection{Quantum phases vis-a-vis
quantum critical points}
\label{phases}

The different types of magnetic QCPs, illustrated in Fig.~\ref{theory},
suggest that the magnetic quantum phases also come in different classes.
The presence or absence of Kondo screening, which influences quantum
criticality, also leads to distinct Fermi surfaces in a magnetically
ordered
HF metal. The issue is of larger importance: The need to use the
Fermi surface to distinguish between various
zero-temperature magnetic phases implies
the failure of the
Landau
assumption that
an
order parameter alone
characterizes
quantum phases.

The Kondo
screening
deep inside the AF part of the Kondo-lattice phase
diagram was studied recently.\cite{Yamamoto07,Si06} The Kondo
interaction is shown to be exactly marginal in the renormalization
group sense
(with respect to the
zero-Kondo-coupling fixed point). The result provides a firm
theoretical ground for the existence of an AF metal with a small
Fermi surface -- an ${\rm AF_S}$ phase in which the local moments
(behaving as localized core electrons) do not participate in the
formation of the electronic Fermi volume. This phase is different
from the ${\rm AF_L}$ phase, in which the Fermi surface is large
in the sense that the local moments are part of the electronic
fluid.
Note that, because the AF ordering breaks the lattice translational
symmetry, the Brillouin zone in a commensurate antiferromagnet is
half that of the paramagnetic state. When counted in the magnetic
Brillouin zone, the distinction between the ${\rm AF_S}$ and ${\rm
AF_L}$ phases does not lie in the Fermi volume, but in the
Fermi-surface topology. The two phases are separated by a Lifshitz
transition. The same applies to incommensurate antiferromagnets, to
linear order in the magnetic order parameter.
Note also that the ${\rm AF_S}$ and ${\rm AF_L}$
phases may share the same magnetic structure;
the main distinction between the two occurs in the electronic
excitations, {\it i.e.} in the nature of Fermi
surfaces.

In the paramagnetic phase, on the other hand, the Kondo interaction
is traditionally understood to be relevant in the renormalization
group sense.\cite{Hewson}
The
strong-coupling nature of the
fixed point signifies that the local moments and spins of the
conduction electrons form a (Kondo) singlet ground state.
As described in Sec.~\ref{heavy_fermion_intro},
the Kondo resonances develop and the Fermi surface is large
in this paramagnetic metal phase
(${\rm PM_L}$). \cite{Hewson}


There has been a long history of experimental studies of the Fermi
surface at low temperatures by probing
dHvA
oscillations (for a recent review, see Ref. \cite{Onuki}).
Measurements in the antiferromagnetically ordered HF metals
have provided extensive evidence for the ${\rm AF_S}$
phase; a typical example is ${\rm CeRh_2Si_2}$, which is
antiferromagnetically ordered at ambient pressure. In paramagnetic
HF metals, evidence
for the ${\rm PM_L}$ phase has existed since the early days
of
HF studies.


One intriguing aspect with the ${\rm AF_S}$ phase concerns the
quasiparticle mass. Mass enhancement in HF metals is traditionally
associated with the Kondo resonances, as is appropriate for both the
${\rm PM_L}$ and ${\rm AF_L}$ phases. How can a heavy mass arise in
the ${\rm AF_S}$ phase? To see this, it is important to recognize
that the Kondo destruction in the ${\rm AF_S}$ phase refers to the
vanishing of the {\it static} amplitude of the Kondo singlet.
The small Fermi surface
is a property of the ground state;
in other words, the discontinuity in the ground-state
momentum distribution occurs at the small Fermi surface,
and
not
where the large Fermi surface would be located. The effective mass,
on the other hand, is a dynamical quantity; it measures the
dispersion of single-electron {\it excitations}, that is, quasiparticles.
The effective mass in the
${\rm AF_S}$ phase is enhanced due to the {\it dynamical} part of
the Kondo-singlet correlations. That dynamical Kondo correlations
should persist in the ${\rm AF_S}$ phase is intuitively clear. It
has been demonstrated in some dynamical studies of the magnetically
ordered region of the Kondo lattice
model,\cite{Zhu_03,Glossop07,Zhu07} and is also in part reflected in
the marginal nature of the Kondo coupling in the renormalization-group
study mentioned
earlier.\cite{Yamamoto07,Si06}

The mass enhancement in the ${\rm AF_S}$ phase is also important for
the Kondo-destroying QCP. Since the effective mass diverges as the
QCP is approached from the paramagnetic side, continuity dictates
the same to be true as the QCP is approached from the magnetically
ordered side. Experimentally, a large effective mass has been
observed in the ordered states of both CeCu$_{6-x}$Au$_x$ and
YbRh$_2$Si$_2$. \cite{Loehneysen,Gegenwart 2002}

We close this Subsection by addressing the theoretical studies of
the destruction of the Kondo effect within single-impurity Kondo-type
problems. These works complement the studies already mentioned in
the case of Kondo lattice systems;
through extended dynamical mean field studies,
for instance, these works
have had considerable crosstalk with
those on the lattice problem.
The physics of the critical Kondo-destruction effect and
the associated scaling properties have been most extensively studied
within the Bose-Fermi Kondo
model.\cite{Zhu02,Zarand02,Vojta03,Glossop05,Kirchner07}
A related Kondo-destruction effect has also been studied in a
single-impurity Kondo problem whose conduction-electron host is
placed near its spin-density-wave QCP.\cite{Maebashi}
Finally, the formulation of the
single-impurity Kondo problem in a Schwinger-boson
representation,\cite{Rech06} while not directly concerned with the
Kondo-destruction effect, holds the promise to handle magnetism and
Kondo effect of the Kondo-lattice problem within a large-N
treatment.

\subsection{Other potentially quantum critical heavy fermion metals}
\label{potential_qcp}

From the materials perspective, the quantum critical
HF metals we have most extensively discussed so far ({\it cf.}
Fig.~\ref{overall}) are distinct in that the existence of a QCP has
been established explicitly: AF order has been observed and,
moreover, has been continuously suppressed to $T=0$.
There is a sizable number of other
HF metals in which QCPs have been implicated.

One case is ${\rm
CeRhIn_5}$, whose low-temperature phase diagram is gradually being
uncovered. It is an antiferromagnet at ambient pressure and turns
into a superconductor when the applied pressure is sufficiently
large. Applying a magnetic field inside the superconducting region
induces a smooth transition between a purely superconducting phase
and one in which superconductivity and antiferromagnetism
co-exist.\cite{Park_06,Knebel_06} A particularly exciting
possibility occurs at a magnetic field of $>9$~T, where applying
pressure induces a direct second-order transition from a purely AF
phase to a non-magnetic one. This has been hinted through an
extrapolation of the lower-field measurements of
Ref.~\cite{Park_06}. However, direct thermodynamic/transport/neutron
scattering measurements are not yet available.

\begin{figure}[!ht]
\begin{center}
\vskip 0.5 cm
\includegraphics[angle=0,width=0.75\textwidth]{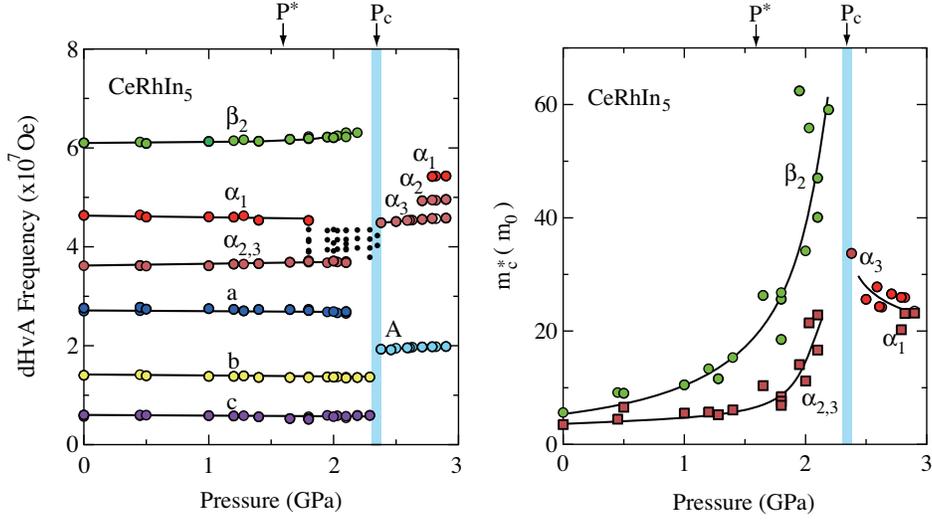}
\end{center}
\caption{Changes of Fermi surface properties across a likely quantum
critical point in CeRhIn$_5$. The left and right panels show the
pressure dependence of the de Haas-van Alphen frequencies and
cyclotron mass, respectively.\cite{Shishido}
}\label{fig8}
\end{figure}

Establishing such a high-field, high-pressure AF QCP is especially
important in light of the existing dHvA results, \cite{Shishido}
which
were obtained in
magnetic fields of the order of 10~T.
Fig.~\ref{fig8} illustrates the potential jump of the Fermi volume,
and divergence of the cyclotron mass, at a pressure coinciding with
the critical pressure ${\rm p_c} \approx 2.3$~GPa for the
putative
QCP
({\it cf.} our discussion on CeRu$_2$Si$_2$ in subsection VII E).
These observations imply
that, if the underlying physics is indeed
a second-order AF quantum phase
transition, the quantum criticality would be of the Kondo-destroying
type. By extension,
it could be that the nearby superconductivity is
promoted by unconventional quantum criticality.

The situation in ${\rm CeCoIn_5}$ is more complex.
There have
been extensive measurements which have implicated an AF QCP slightly
below the upper critical field for
superconductivity.\cite{LANL,Sherbrooke, Wirth} Results of thermal
expansion measurements on ${\rm CeCoIn_5}$, performed as a function
of temperature at magnetic fields slightly larger than ${\rm
H_{c2}}(0)$, suggest a crossover from 2D to 3D quantum critical
spin-density-wave fluctuations when the sample is cooled to below $T
\approx 0.3$~K. \cite{Donath} While long-range AF order has not yet
been observed in the pure compound, it was recently
identified
in ${\rm CeCoIn_5}$ moderately doped with Cd. \cite{Pham_06} It will
be important to establish whether the AF order can be continuously
suppressed as a function of Cd-doping, perhaps in high magnetic
fields.

Pronounced NFL behavior in thermodynamic and transport properties
have also been found in the normal-state of the actinide HF
superconductor UBe$_{13}$ and tentatively been ascribed to a
field-induced AF QCP
covered by the superconducting state. \cite{Gegenwart_04}

${\rm CeNi_2Ge_2}$ is a prototypical HF metal showing NFL phenomena
in accord with a 3D spin-density-wave QCP scenario. \cite{Kuechler
2003} Under increasing magnetic field, the crossover temperature
from the NFL regime to the Landau FL one was found to continuously
increase. \cite{Gegenwart_99} As in the case of ${\rm CeCoIn_5}$, AF
order does not occur in the pure ${\rm CeNi_2Ge_2}$ but
has been seen when Ni is substituted by,
{\it eg.}, Cu (Ref. \cite{Loidl 92}).

In cubic ${\rm CeIn_3}$, AF order ($T_N \approx 10$~K) can be
continuously suppressed by substituting Sn for In.
The data are consistent with
a QCP of the 3D
spin-density-wave type at $x_c \approx 0.67$ and a low-temperature
Landau FL phase at $x > x_c$. \cite{Kuechler 2006}
The extent to which disorder influences
the quantum critical properties of this material remains to be
clarified. 

Inelastic neutron scattering studies have played a vital role in our
understanding of quantum criticality in HF
metals. Indeed, $\omega/T$ scaling in the inelastic neutron scattering
spectrum was observed very early on, in
the Pd-doped ${\rm U Cu_5}$ system,\cite{Aronson}, and
was subsequently studied in the Au-doped CeCu$_{6}$
system.\cite{Schroeder,Stockert}
Single
crystals of the
HF system Ce(Ru${\rm _{1-x}}$Rh$_{\rm x}$)$_{\rm 2}$Si${\rm _2}$
show
an AF ordering
for relatively small $x$.
The thermodynamic or transport
data are consistent with a magnetic QCP
at $x_c \approx 0.04$, although the phase
boundary has not been followed
to low enough temperatures (below 3 K).
Near $x_c$, the inelastic neutron
scattering data~\cite{Kadowaki}
are
well described by the
form, $\chi({\bf q},\omega ) = {\rm \chi_{\bf q}(T)}/
{[1-i\omega / \Gamma_{\bf q}(T)]}$, with $\Gamma_{\bf Q} \sim
T^{3/2}$.
The violation of $\omega/T$ scaling occurs in a way that it
is consistent
with the 3D spin-density-wave QCP.\cite{Hertz,Moriya,Millis}
The magnetic fluctuations are three-dimensional, \cite{Kadowaki} in
contrast to the quasi-2D behavior~\cite{Stockert,Schroeder} seen in
CeCu${\rm _{6-x}}$Au${\rm _x}$.
Separately, neutron scattering results on Ce$_{1-x}$La$_x$Ru$_2$Si$_2$
indicate that
it displays a 3D spin-density-wave QCP with a critical concentration
$x_c\approx 0.075$. 
\cite{Knafo 04}
Those on Ce(Ru$_{1-x}$Fe$_x$)$_2$Ge$_2$ have revealed some complex
dynamical scaling behavior around $x_c\approx 0.77$.
\cite{Montfrooij03}

\subsection{Generality of multiple energy scales}

The transport and thermodynamic measurements of the multiple energy
scales have considerably
clarified the
quantum
criticality in YbRh$_2$Si$_2$. It is natural to ask about the extent
to which the multiple energy scales characterize other
HF
metals
near their QCPs.

As described earlier, the quantum-critical properties of Au-doped
CeCu$_{6}$ share many similarities with those of YbRh$_2$Si$_2$.
Together, the two materials represent the
clearest
candidates for
Kondo-destroying quantum criticality. It will be important to
determine whether a $T^*$ line
meeting
the $T_N$ and
$T_{\text{\small FL}}$
lines in the zero-temperature limit
(Fig.~\ref{YbRh$_2$Si$_2$_overall_phase_diagram}, for
YbRh$_2$Si$_2$)
also occurs in CeCu$_{\rm
6-x}$Au$_{\rm x}$. At $x=x_c=0.1$, this can be done simply by studying
the isothermal field-dependence of various thermodynamic and transport
properties.

At $x>x_c$, such studies can clarify the interesting
contrast between the pressure-induced and magnetic-field-induced
QCPs. The pressure-induced QCP
at $x>x_c$ shares the properties of the QCP
in the $x=x_c$ compound at ambient pressure.\cite{Loehneysen_07}
The field-induced QCP at $x>x_c$, on the other hand,
behaves very differently.
Such a field-induced QCP was first realized
in the system CeCu$_{6-x}$Ag$_x$ ($x_c \approx 0.8$),
with indications for
a field-induced
3D spin-density-wave QCP;\cite{Heuser 98b}
this behavior is in sharp contrast to the
zero-field quantum critical behavior observed in
CeCu$_{5.9}$Au$_{0.1}$. \cite{Schroeder,Stockert}
The inelastic neutron scattering results at the field-induced QCP
in CeCu$_{5.8}$Au$_{0.2}$ are also compatible with a QCP of
the 3D spin-density-wave type.\cite{Stockert Enderle}
Theoretical understanding of this contrast between the
pressure- and field-induced QCPs is not yet available.
Studying the $T^*$ crossover line
should shed much new light on this and related issues.

Even in the case of YbRh$_2$Si$_2$, it will be important to understand
how the relationship between the $T^*$, $T_N$ and
$T_{\text{\small FL}}$
lines evolve when the strength of magnetic ordering is modified,
either by Co- and Ir- doping
[Westerkamp, T. {\it et al.}
{\it Physica B}, in press (2007)]
or by applying pressure\cite{Trovarelli}.

We have also mentioned a number of candidate
HF metals in which the QCPs may be of the 3D spin-density-wave type,
in particular CeCu$_{\rm 2}$Si$_{\rm 2}$.
These cases also lead to
several interesting questions.
As suggested by
Fig.~\ref{theory}b, the
$T^*$ crossover line can in
principle
be located
not too far away from
the magnetic QCP.
Moreover, as discussed in Subsection~\ref{phases}, in the
zero-temperature limit, Fig.~\ref{theory}b implies a Lifshitz
transition within the AF part of the phase diagram. Its existence
will also have to be checked for experimentally.

\subsection{Heavy fermion metals versus weak itinerant antiferromagnets}

Our focus in this article has been on the
AF HF metals.
The Kondo
phenomenon
is central to the
HF
physics {\it per se}.
Moreover, since the Kondo
effect
involves quantum entanglement,
critical Kondo destruction
naturally introduces
inherently quantum excitations that are beyond those associated
with the slow fluctuations of the AF order parameter.

Given this emphasis on the Kondo
phenomenon,
it will be important to
compare and contrast the quantum criticality of AF
HF metals with that of transition metal-based
weak itinerant antiferromagnets. In the
latter systems,
the Kondo effect is
not expected to play any role
(since, here,
the $d$-electrons are itinerant - in contrast
to the localized $f$-electrons)
and it would be more
natural to apply the spin-density-wave QCP picture. Unfortunately,
the situation is not so straightforward. There are issues both
experimentally and theoretically.

Experimentally, a prototype example for a second-order quantum phase
transition in weak itinerant antiferromagnets is V-doped Cr.
We have already mentioned the signatures of
the magnetic quantum phase transition in Hall-coefficient
measurements. Transport and thermodynamic experiments, however, have
failed to see any signature of the NFL excitations. Presumably, the
reason has to do with the relative weakness of the electron-electron
interactions, which makes the amplitude of the NFL parts in, {\it
e.g.} the electrical resistivity and specific heat,
relatively
small. Better candidate materials to study the NFL excitations in
the quantum critical regime of
weak itinerant antiferromagnets
will probably come from transition-metal-based intermetallic
compounds.

Very recently, a logarithmic divergence of the low-temperature
specific heat coefficient and a $\Delta\rho\propto T^{1.5}$
dependence in the electrical resistivity has been observed in a
slightly Nb-rich single crystal of the d-electron system NbFe$_2$
[Brando, M. {\it et al.}, to be published].
In this case, the fluctuation spectrum may be
complicated since strong {\it ferromagnetic} fluctuations seem to be
present close to the disappearance of a (possibly long-wave length)
spin-density-wave phase.

From a theoretical perspective, even though the Kondo effect is
absent, these systems still contain gapless electronic excitations.
Studies by various groups have shown that in the process of
integrating out such gapless electronic excitations and constructing
an effective theory for the order-parameter fluctuations,
non-analyticity develops in the coupling constants for the latter,
which may in turn change the form of the quantum criticality.
\cite{Belitz_rmp05,Chubukov}

\subsection{The case of ferromagnetic quantum criticality}

Ferromagnetic quantum phase transitions have hardly been studied in
HF metals. Ferromagnetic ordering in HF metals is rare to begin
with, compared to AF ordering. What is more, in cases in which the
Curie temperature can be tuned, other instabilities -- such as those
leading to AF phases -- typically occur when the Curie temperature is
sufficiently reduced. \cite{Sullow_prl99,Fisk-ferro}

In the weak itinerant ferromagnets \cite{Lonzarich} such as
MnSi,\cite{Lonzarich03}
ZrZn$_{\rm 2}$ \cite{ZrZn2} and FeGe, \cite{FeGe}
the transition
turns into first order when the Curie temperature is sufficiently
reduced.
(More precisely, MnSi displays a spiral magnetic ordering with a long
spiral pitch.)
One of the outstanding questions is why NFL behavior still
occurs even though the quantum phase transition itself is of
first-order.
In fact, in MnSi, the NFL behavior in the electrical
resistivity occurs in such a wide pressure range (from the critical
pressure $p_c$ up to at least $2p_c$)\cite{Lonzarich03}
that it may very well be the property of a NFL {\it phase}, as
opposed to being associated with
the fluctuations of any quantum critical {\it point}.

Theoretically, the non-analyticities associated with integrating out
the gapless conduction electrons of the weak itinerant ferromagnets
provide a generic mechanism for the first order nature of the
ferromagnetic quantum phase transitions.\cite{Belitz_rmp05} In the
presence of Ising anisotropy, however, such non-analyticity effects
should not occur,\cite{Belitz_rmp05} and the corresponding quantum
criticality is supposed to be of the order-parameter-fluctuation
type. This makes it particularly appealing to study quantum
criticality in metamagnetic systems.

A metamagnetic transition describes the sudden onset of the
magnetization as a function of magnetic field. This transition
involves two phases of the same symmetry.
Being in the same
class as the melting transition of
ice to water, it is of first order.
A line of first order transitions
in this case
will terminate at a critical end point. By tuning
one additional parameter,
the critical end point can
be driven to
zero
temperature. \cite{Schofield-Millis} In that case, a {\it quantum}
critical end point arises. The most systematic studies have been
carried out in recent years in the metallic bilayer ruthenate
Sr$_3$Ru$_2$O$_7$. \cite{grigera-science01} Here, the field-tuned
QCP is masked by a new, as yet
unidentified, low-temperature state.
\cite{Gegenwart Sr327,grigera-science04}
In Ref. \cite{Flouquet02}, Sr$_3$Ru$_2$O$_7$ has been compared with
the prototypical metamagnetic HF compound CeRu$_2$Si$_2$.
Transport and thermodynamic measurements\cite{Weickert},
however, have proved a Landau FL state in the latter.
This observation is
consistent with a continuous evolution of the Fermi surface
across the metamagnetic transition as inferred from recent
Hall-effect and dHvA investigations. \cite{julian06}

The related U-based heavy fermion compound URu$_2$Si$_2$ shows
metamagnetism around 38~T.\cite{Kim 2003}
However, in contrast to CeRu$_2$Si$_2$,
the low-temperature magnetization indicates three subsequent
first-order metamagnetic transitions, as the magnetic field
is increased.
The first transition is related to the suppression of the as yet
unidentified ``hidden order'' (whose zero-field transition temperature
is $T_0=17.5$~K), whereas the second and third transition points
bound a phase which has been labeled  ``re-entrant hidden
order'';
transport experiments suggest that the latter phase hides a
metamagnetic quantum critical end point.\cite{Kim 2003}



\section{\bf Summary and Outlook}


The explicit observation of
antiferromagnetic
quantum critical points in
heavy fermion
metals
has provided the opportunity to assess the role quantum criticality
plays in the general physics of strongly correlated electron
systems. Through the studies of quantum critical
heavy fermion
metals, we now
know that quantum criticality
indeed generates
non-Fermi liquid
excitations, and does so in a robust fashion.
We also know that
it can lead to unconventional superconductivity.

Studies in
heavy fermion
metals have also elucidated the nature of quantum
criticality. We now have clear evidence that quantum criticality can
go beyond the standard theory of order-parameter fluctuations. The
destruction of Kondo entanglement has emerged as a means of
generating
the inherently quantum excitations that
add
to the
critical fluctuations of the order parameter. The evidence for such
``Kondo collapsing'' at the quantum critical point appears in the form
of unconventional scalings, a divergence of the effective
charge-carrier mass, a sudden reconstruction of the Fermi surface,
and the vanishing of multiple energy scales.

Still, the subject of quantum critical
heavy fermion
metals is at a very early
stage. Many important questions remain, and we list several here to
close this review.

\begin{itemize}

\item We have discussed two types of quantum criticality in
antiferromagnetic
heavy fermion
metals, one involving critical Kondo destruction and the
other one being of the spin-density-wave type. Do all
antiferromagnetic
heavy fermion
quantum critical points fall into one of these two categories?

\item In classical critical phenomena, the identification of the
critical modes -- fluctuations of the order parameter -- immediately
leads to a field theory, the Ginzburg-Landau theory.
The construction of this field theory is important for the concept
of universality.
For the spin-density-wave quantum critical points,
the procedure is readily adapted.
With the new
excitations from the
critical Kondo destruction,
on the other hand,
what is the form
of the field theory?

\item We have outlined the empirical evidence that the spin-density-wave
type of quantum criticality promotes unconventional superconductivity.
None
of the materials with the clearest evidence for Kondo-destroying
quantum criticality
have
shown superconductivity. Is this an indication
of a clear-cut correlation between the type of quantum criticality
and the occurrence of superconductivity?

\item What lessons
can be learnt from
heavy fermion
metals that are
directly relevant to the quantum criticality of other strongly
correlated systems?

\end{itemize}

\vspace{0.5cm} \noindent{\bf Acknowledgments} We would like to thank
E. Abrahams,
C. J. Bolech, M. T. Glossop, S. Kirchner, M. Nicklas, P. Nikolic,
S. Paschen, T. Westerkamp, S. Wirth, and S. J. Yamamoto
for their helpful input on
the manuscript. The work at G\"{o}ttingen and Dresden has been
supported in part by the DFG Research Unit 960 ("Quantum Phase
Transitions") and the work at Rice by NSF Grant No. DMR-0706625 and
the Robert A. Welch Foundation.

\vspace{0.5cm} \noindent{\bf Competing interests statement} The
authors declare that they have no competing financial interests.

\vspace{0.5cm} \noindent{\bf Correspondence} and requests for
materials should be addressed to Q.~S. (qmsi@rice.edu).


\begin{thebibliography}{99}

\bibitem{ice} Rosenberg, R.
Why is ice slippery? {\it Phys. Today} {\bf 58} (12), 50-55 (2005).

\bibitem{Sachdev_book} Sachdev, S. {\it Quantum Phase Transitions}
(Cambridge Univ. Press, New York, 1999).

\bibitem{Coleman-Schofield} Coleman, P. \& Schofield, A.J. Quantum criticality. {\it Nature} {\bf 433},
226-229 (2005).

\bibitem{Stewart_01} Stewart, G.R. Non-Fermi-liquid behavior in d- and f-electron metals. {\it Rev. Mod. Phys.} {\bf 73},
797-855 (2001), {\bf 78}, 743-753 (2006).

\bibitem{Loehneysen_07} L\"{o}hneysen, H.v., Rosch, A., Vojta, M. \& W\"{o}lfle, P. Fermi-liquid instabilities at magnetic quantum phase
transitions. {\it Rev. Mod. Phys.} {\bf 79}, 1015-1075 (2007).

\bibitem{Loehneysen} L\"{o}hneysen, H.v. {\it et al.} Non-Fermi-liquid behavior in a heavy-fermion alloy at a magnetic
instability. {\it Phys. Rev. Lett.} {\bf 72}, 3262-3265 (1994).

\bibitem{Schroeder} Schr\"{o}der, A. {\it et al.} Onset of antiferromagnetism in heavy-fermion metals. {\it Nature} {\bf 407}, 351-355 (2000).


\bibitem{Trovarelli} Trovarelli, O. {\it et al.} YbRh$_2$Si$_2$: Pronounced non-Fermi-liquid effects above a low-lying magnetic phase transition. {\it Phys. Rev.
Lett.} {\bf 85}, 626-629 (2000).


\bibitem{Custers} Custers, J. {\it et al.} The break-up of heavy electrons at a quantum critical point. {\it Nature} {\bf 424}, 524-527 (2003).



\bibitem{Mathur} Mathur, N.D. {\it et al.} Magnetically mediated superconductivity in heavy fermion compounds. {\it Nature} {\bf 394}, 39-43 (1998).

\bibitem{Hertz} Hertz, J.A. Quantum critical phenomena. {\it Phys. Rev. B} {\bf 14}, 1165-1184 (1976).

\bibitem{Millis} Millis, A.J. Effect of a nonzero temperature on quantum critical points in
itinerant fermion systems. {\it Phys. Rev.} {\bf B 48}, 7183-7196
(1993).

\bibitem{Moriya} Moriya, T. \& Takimoto, T. Anomalous properties around magnetic instability in heavy electron systems. {\it J. Phys. Soc. Jpn.} {\bf 64}, 960-969 (1995).

\bibitem{Si} Si, Q., Rabello, S., Ingersent, K. \& Smith, J.L. Locally critical quantum phase transitions
in strongly correlated metals. {\it Nature} {\bf 413}, 804-808
(2001).

\bibitem{Coleman} Coleman, P., P\'{e}pin, C., Si, Q. \& Ramazashvili,
R. How do Fermi liquids get heavy and die? {\it J. Phys.: Cond.
Matt.} {\bf 13}, R723-R738 (2001).

\bibitem{Senthil} Senthil, T., Vojta, M. \& Sachdev, S. Weak magnetism and non-Fermi liquids
near heavy-fermion critical points. {\it Phys. Rev. B} {\bf 69},
035111 (2004).


\bibitem{Andres} Andres, K., Graebner, J.E. \& Ott, H.R.
4f-virtual-bound-state formation in CeAl$_3$ at low temperatures.
{\it Phys. Rev. Lett.} {\bf 35}, 1779-1782 (1975).

\bibitem{Steglich 79} Steglich, F. {\it et al.} Superconductivity in the presence of strong Pauli paramagnetism:
CeCu$_2$Si$_2$. {\it Phys. Rev. Lett.} {\bf 43}, 1892-1896 (1979).

\bibitem{Assmus} Assmus, W. {\it et al.} Superconductivity in CeCu$_2$Si$_2$ single
crystals. {\it Phys. Rev. Lett.} {\bf 52}, 469-472 (1984).

\bibitem{Ott 83} Ott, H.R., Rudigier, H., Fisk, Z. \&
Smith, J.L. UBe$_{13}$: An unconventional actinide superconductor.
{\it Phys. Rev. Lett.} {\bf 50}, 1595-1598 (1983).

\bibitem{Stewart UPt3} Stewart, G.R., Fisk, Z., Willis, J.O.
\& Smith, J.L. Possibility of coexistence
 of bulk superconductivity and spin fluctuations in UPt$_3$.
{\it Phys. Rev. Lett.} {\bf 52}, 679-682 (1984).

\bibitem{Schlabitz} Schlabitz, W. {\it et al.} Superconductivity and
magnetic order in a strongly interacting Fermi system:
URu$_2$Si$_2$. Abstract ICVF, Cologne, August 1984 (unpublished);
{\it Z. Phys. B} {\bf 62}, 171-177 (1986).

\bibitem{Grewe} Grewe, N. \& Steglich, F.
Heavy fermions. {\it Handbook on the physics and chemistry of rare
earths} {\bf 14}, Gschneidner Jr., K.~A. \& Eyring, L., eds.
(Elsevier, Amsterdam, 1991), p. 343-474, and references therein.




\bibitem{Kondo}
Kondo, J. Resistance minimum in dilute magnetic alloys. {\it Prog.
Theor. Phys.} {\bf 32}, 37-49 (1964).

\bibitem{Hewson} Hewson, A.C. {\it The Kondo problem to heavy
fermions} (Cambridge University Press, Cambridge, England, 1993).













\bibitem{KotliarVollhardt}
Kotliar, G. \& Vollhardt, D. Strongly correlated materials: Insights
from dynamical mean-field theory. {\it Physics Today} {\bf 57 (3)},
53-59 (2004).

\bibitem{Doniach} Doniach, S. The Kondo lattice and weak antiferromagnetism. {\it Physica B+C} {\bf 91}, 231-234 (1977).

\bibitem{Varma76} Varma, C.M. Mixed-valence compounds. {\it Rev.
Mod. Phys.} {\bf 48}, 219-238 (1976).


\bibitem{Seaman 91} Seaman, C.L. {\it et al.} Evidence for non-Fermi liquid behavior in the Kondo alloy
Y$_{1-x}$U$_x$Pd$_3$. {\it Phys. Rev. Lett.} {\bf 67}, 2882-2885
(1991).

\bibitem{Andraka 91} Andraka, B. \& Tsvelik, A.M. Observation of non-Fermi-liquid behavior in
U$_{0.2}$Y$_{0.8}$Pd$_3$. {\it Phys. Rev. Lett.} {\bf 67}, 2886-2889
(1991).



\bibitem{Chakravarty_89}
Chakravarty, S., Halperin, B.\ I.\ \& Nelson, D.\ R. Two-dimensional
quantum Heisenberg antiferromagnet at low temperatures. {\it Phys.\
Rev.\ B} {\bf 39}, 2344-2371 (1989).

\bibitem{Continentino}
Continentino, M.\ A., Japiassu, G.\ M. \&Troper, A. Critical
approach to the coherence transition in Kondo lattices.
{\it Phys.\ Rev.\ B} {\bf 39}, 9734-9737 (1993).


\bibitem{Senthil_deconfined} Senthil, T., Vishwanath, A., Balents, L., Sachdev, S. \& Fisher, M.P.A. Deconfined quantum critical points. {\it Science} {\bf 303}, 1490-1494 (2004).

\bibitem{Si_03}
Si, Q., Rabello, S., Ingersent, K. \& Smith, J.L. Local fluctuations
in quantum critical metals. {\it Phys. Rev.\ B} {\bf 68}, 115103
(2003).

\bibitem{Georges}
Burdin, S., Grempel, D.~R. \& Georges, A. Heavy-fermion and
spin-liquid behavior in a Kondo lattice with magnetic frustration.
{\it Phys.\ Rev.\ B} {\bf 66}, 045111 (2002).


\bibitem{FS} Steglich, F. Magnetic moments of rare earth ions in a
metallic environment. Festk\"{o}rperprobleme (Adv. Solid State
Physics), Treusch, J., ed., Vol XVII (Vieweg, Braunschweig 1977), p.
319-350.

\bibitem{Melnikov} Melnikov, V.I.. Thermodynamics of the Kondo problem. {\it Sov. Phys. JETP Lett.}, {\bf
35}, 511-515 (1982).


\bibitem{Zhu_03}
Zhu, J.-X., Grempel, D.~R. \& Si, Q. Continuous quantum phase
transition in a Kondo lattice model. {\it Phys. Rev. Lett.} {\bf
91}, 156404 (2003).

\bibitem{Glossop07}
Glossop, M.T. \& Ingersent, K. Magnetic quantum phase transition in
an anisotropic Kondo lattice. {\it Phys. Rev. Lett.} {\bf 99},
227203 (2007).

\bibitem{Zhu07}
Zhu, J.-X., Kirchner, S., Bulla, R. \& Si, Q. Zero-temperature
magnetic transition in an easy-axis Kondo lattice model, {\it Phys.
Rev. Lett.} {\bf 99}, 227204 (2007).


\bibitem{zhu} Zhu, L., Garst, M., Rosch, A. \& Si, Q. Universally diverging Gr\"{u}neisen parameter and
the magnetocaloric effect close to quantum critical points. {\it
Phys. Rev. Lett.} {\bf 91}, 066404 (2003).

\bibitem{garst} Garst, M. \& Rosch, A. Sign change of the Gr\"{u}neisen parameter and
magnetocaloric effect near quantum critical points. {\it Phys. Rev.
B} {\bf 72}, 205129 (2005).

\bibitem{Kuechler 2003} K\"{u}chler, R. {\it et al.} Divergence of the Gr\"{u}neisen
ratio at quantum critical points in heavy fermion metals. {\it Phys.
Rev. Lett.}~{\bf 91}, 066405 (2003).

\bibitem{Kuechler 2006}
K\"{u}chler, R. {\it et al.} Quantum criticality in the cubic
heavy-fermion system CeIn$_{3-x}$Sn$_x$. {\it Phys. Rev. Lett.}~{\bf
96}, 256403 (2006).

\bibitem{Gegenwart Sr327} Gegenwart, P., Weickert, F., Garst, M., Perry, R.S. \& Maeno, Y. Metamagnetic
 quantum criticality in Sr$_3$Ru$_2$O$_7$ studied by thermal expansion. {\it Phys. Rev. Lett.} {\bf 96}, 136402 (2006).


\bibitem{Gegenwart 2007} Gegenwart, P. {\it et al.} Multiple energy scales at a quantum critical point.
{\it Science} {\bf 315}, 969-971 (2007).


\bibitem{Pepin05}
P\'{e}pin, C. Fractionalization and Fermi-surface volume in
heavy-fermion compounds: The case of YbRh$_2$Si$_2$. {\it Phys. Rev.
Lett.} {\bf 94}, 066402 (2005).

\bibitem{Kuechler 2004} K\"{u}chler, R. {\it et al.}. Gr\"{u}neisen
ratio divergence at the quantum critical point in
CeCu$_{6-x}$Ag$_x$. {\it Phys. Rev. Lett.}~{\bf 93}, 096402 (2004).

\bibitem{Varma_MFL}
Varma, C.\ M., Littlewood, P.\ B., Schmitt-Rink, S., Abrahams, E.\
\& Ruckenstein, A.\ E. Phenomenology of the normal state of Cu-O
high temperature superconductors. {\it Phys.\ Rev.\ Lett.} {\bf 63},
1996-1999 (1989).

\bibitem{Aronson}
Aronson, M.C. {\it et al.} Non-Fermi-liquid scaling of the magnetic
response in UCu$_{5-x}$Pd$_x$($x=1,1.5$). {\it Phys.\ Rev.\ Lett.}
{\bf 75}, 725-728 (1995).

\bibitem{Stockert} Stockert, O., L\"{o}hneysen, H.v., Rosch, A., Pyka,
N. \& Loewenhaupt, M. Two-dimensional fluctuations at the quantum
critical point of CeCu$_{6-x}$Au$_x$. {\it Phys. Rev. Lett.} {\bf
80}, 5627-5630 (1998).

\bibitem{Geg 05} Gegenwart, P., Custers, J., Tokiwa, Y., Geibel, C. \& Steglich, F. Ferromagnetic
quantum critical fluctuations in
YbRh$_2$(Si$_{0.95}$Ge$_{0.05}$)$_2$. {\it Phys. Rev. Lett.} {\bf
94}, 076402 (2005).


\bibitem{Imada}
Misawa, T., Yamaji, Y. \& Imada, M. YbRh$_2$Si$_2$: Quantum
tricritical behavior in itinerant electron systems. arXiv:0710.3260.

\bibitem{Sichelschmidt} Sichelschmidt, J., Ivanshin, V.A., Ferstl, J., Geibel, C. \&
Steglich, F. Low temperature electron spin resonance of the Kondo
ion in a heavy fermion metal: YbRh$_2$Si$_2$. {\it Phys. Rev. Lett.}
{\bf 91}, 156401 (2003).


\bibitem{Kadowaki} Kadowaki, H. {\it et al.} Quantum critical point of an itinerant antiferromagnet in a heavy
fermion. {\it Phys. Rev. Lett.} {\bf 96}, 016401 (2006).


\bibitem{Gegenwart 1998} Gegenwart, P. {\it et al.} Breakup of heavy fermions on the brink of "phase A" in CeCu$_2$Si$_2$. {\it Phys. Rev. Lett.} {\bf 81},
1501-1504 (1998).



\bibitem{Hlubina} Hlubina, R. \& Rice, T.M. Resistivity as a function of
 temperature for models with hot spots on the Fermi surface. {\it Phys. Rev. B} {\bf 51},
 9253-9260 (1995).

\bibitem{Rosch} Rosch, A. Interplay of disorder and spin fluctuations in the resistivity near a quantum critical
point. {\it Phys. Rev. Lett.} {\bf 82}, 4280-4283 (1999).




\bibitem{Sereni 97} Sereni, J. {\it et al.} Scaling the $C\propto T\ln T$ dependence in Ce-systems. {\it Physica B} {\bf 230-232}, 580-582 (1997).


\bibitem{Paschen} Paschen, S. {\it et al.} Hall-effect evolution across a heavy-fermion quantum critical point. {\it Nature} {\bf 432}, 881-885 (2004).



\bibitem{Norman_prl03} Norman, M.R., Si, Q., Bazaliy, Ya.B. \& Ramazashvili, R. Hall
 effect in nested antiferromagnets near the quantum critical point. {\it Phys.
Rev. Lett.} {\bf 90}, 116601 (2003).

\bibitem{Lee_prl05} Lee, M., Husmann, A., Rosenbaum, T.F. \& Aeppli, G. High resolution study of magnetic ordering at absolute
zero. {\it Phys. Rev. Lett.} {\bf 92}, 187201 (2004).

\bibitem{Shishido} Shishido, H., Settai, R., Harima, H. \& \^{O}nuki,
Y. A drastic change of the Fermi
 surface at a critical pressure in CeRhIn$_5$: dHvA study under
 pressure. {\it J. Phys. Soc. Jpn.}~{\bf 74}, 1103-1106 (2005).

\bibitem{Park_06} Park, T. {\it et al.} Hidden magnetism and quantum criticality in the heavy fermion superconductor CeRhIn$_5$. {\it Nature} {\bf 440}, 65-68 (2006).

\bibitem{Tokiwa 2005} Tokiwa, Y. {\it et al.} Field-induced suppression of the heavy-fermion state in YbRh$_2$Si$_2$. {\it Phys. Rev. Lett.} {\bf 94}, 226402 (2005).

\bibitem{Gegenwart 2002} Gegenwart, P. {\it et al.} Magnetic-field induced quantum critical point in YbRh$_2$Si$_2$. {\it Phys. Rev. Lett.} {\bf 89}, 056402 (2002).


\bibitem{Pepin} Paul, I., P\'{e}pin, C. \& Norman, M.~R.
Kondo breakdown and hybridization fluctuations in the
Kondo-Heisenberg lattice. {\it Phys. Rev. Lett.} {\bf 98}, 026402
(2007).

\bibitem{Yeh_nature02} Yeh, A. {\it et al.} Quantum phase transition
in a common metal. {\it Nature} {\bf 419}, 459-462 (2002).

\bibitem{Canfield} Bud'ko, S.L., Morosan, E. \& Canfield, P.C. Magnetic field induced non-Fermi liquid behavior in YbAgGe single crystals. {\it Phys. Rev.
B} {\bf 69}, 014415 (2004).

\bibitem{Canfield2} Bud'ko, S.L., Zapf, V., Morosan, E. \& Canfield, P.C.
Field-dependent Hall effect in single-crystal heavy fermion YbAgGe
below 1~K. {\it Phys. Rev. B} {\bf 72}, 172413 (2005).

\bibitem{Tokiwa YbAgGe} Tokiwa, Y. {\it et al.} Low-temperature
thermodynamic properties of the heavy-fermion compound YbAgGe close
to the field-induced quantum critical point. {\it Phys. Rev. B} {\bf
73}, 094435 (2006).

\bibitem{Miyake}
Onishi, Y. \& Miyake, K. Enhanced valence fluctuations caused by f-c
Coulomb interaction in Ce-based heavy electrons: Possible origin of
pressure-induced enhancement of superconducting transition
temperature in CeCu$_2$Ge$_2$ and related compounds. {\it J. Phys.
Soc. Jpn.} {\bf 69}, 3955-3964 (2000).

\bibitem{Jaccard} Jaccard, D., Behnia, K. \& Sierro, J. Pressure-induced heavy fermion superconductivity in CeCu$_2$Ge$_2$. {\it Phys. Lett. A} {\bf 163}, 475-480 (1992).

\bibitem{Yuan} Yuan, H.Q. {\it et al.} Observation of two distinct superconducting phases in CeCu$_2$Si$_2$. {\it Science} {\bf 302}, 2104-2107 (2003).

\bibitem{Sato01} Sato, N.K. {\it et al.} Strong coupling between
local moments and superconducting 'heavy' electrons in
UPd$_2$Al$_3$. {\it Nature} {\bf 410}, 340-343 (2001).

\bibitem{Saxena} Saxena, S.S. {\it et al.} Superconductivity on the border of itinerant-electron ferromagnetism in UGe$_2$. {\it Nature} {\bf 406}, 587-592 (2000).

\bibitem{Huxley} L\'{e}vy, F., Sheikin, I., Grenier, B. \& Huxley, A.D. Magnetic field-induced superconductivity in the ferromagnet URhGe. {\it Science} {\bf 309}, 1343-1346 (2005).

\bibitem{Huy} Huy, N.T., {\it et al.} Superconductivity on the border of weak itinerant ferromagnetism in
UCoGe. {\it Phys. Rev. Lett.} {\bf 99}, 067006 (2007).


\bibitem{Stockert_04} Stockert, O. {\it et al.} Nature of the A phase in CeCu$_2$Si$_2$. {\it Phys. Rev. Lett.} {\bf 92}, 136401 (2004).




\bibitem{Scalapino_86} Scalapino, D.J., Loh, E. \& Hirsch, J.E. d-wave pairing near a spin-density-wave instability. {\it Phys. Rev. B} {\bf 34}, 8190-8192 (1986).



\bibitem{Petrovic} Petrovic, C. {\it et al.} Heavy-fermion superconductivity in CeCoIn$_5$ at 2.3 K. {\it J. Phys.: Cond. Matter} {\bf 13}, L337-L342 (2001).

\bibitem{Gegenwart_04} Gegenwart, P. {\it et al.} Non-Fermi liquid
normal state of the heavy-fermion superconductor UBe$_{13}$. {\it
Physica C} {\bf 408-410}, 157-160 (2004).





\bibitem{Anderson_07}
Anderson, P.~W. Is there glue in cuprate superconductors? {\it
Science} {\bf 316}, 1705-1707 (2007).


\bibitem{Yamamoto07} Yamamoto, S.J. \& Si, Q. Fermi surface
and antiferromagnetism in the Kondo lattice: An asymptotically exact
solution in $d>1$ dimensions. {\it Phys. Rev. Lett.} {\bf 99},
016401 (2007).

\bibitem{Si06} Si, Q. Global magnetic phase diagram and local quantum criticality in heavy fermion
metals. {\it Physica B} {\bf 378}, 23-27 (2006).


\bibitem{Onuki} Settai, R., Takeuchi, T. \& \^{O}nuki, Y. Recent advances
in Ce-based heavy-fermion superconductivity and Fermi surface
properties. {\it J. Phys. Soc. Jpn.} {\bf 76}, 051003 (2007).












\bibitem{Zhu02}
Zhu., L.\ \& Si, Q. Critical local moment fluctuations in the
Bose-Fermi Kondo model, {\it Phys. Rev. B} {\bf 66}, 024426 (2002).

\bibitem{Zarand02}
Zar\'{a}nd, G.\ \& Demler, E.\ Quantum phase transitions in the
Bose-Fermi Kondo model. {\it Phys.\ Rev.\ B} {\bf 66}, 024427
(2002).

\bibitem{Vojta03}
Vojta, M. \& Kir{\'c}an, M. Pseudogapped Fermi-Bose Kondo model.
{\it Phys. Rev. Lett.} {\bf 90}, 157203 (2003).

\bibitem{Glossop05} Glossop, M.~T. \& Ingersent, K.
Numerical renormalization-group study of the Bose-Fermi Kondo model.
{\it Phys. Rev. Lett.} {\bf 95}, 067202 (2005).

\bibitem{Kirchner07}
Kirchner, S. \& Si, Q. Scaling and enhanced symmetry at the quantum
critical point of the sub-Ohmic Bose-Fermi Kondo model. {\it Phys.
Rev. Lett.} {\bf 100}, 026403 (2008).

\bibitem{Maebashi}
Maebashi, H., Miyake, K. \& Varma, C.~M. Undressing the Kondo effect
near the antiferromagnetic quantum critical point. {\it Phys. Rev.
Lett.} {\bf 95}, 207207 (2005).




\bibitem{Rech06}
Rech, J., Coleman, P., Zar\'{a}nd, G.\ \& Parcollet, O. Schwinger
Boson approach to the fully screened Kondo model, {\it Phys. Rev.
Lett.} {\bf 96}, 016601 (2006).

\bibitem{Knebel_06} Knebel, G., Aoki, D., Braithwaite, D., Salce, B. \& Flouquet, J. Coexistence of
antiferromagnetism and superconductivity in CeRhIn$_5$ under high
pressure and magnetic field. {\it Phys. Rev. B} {\bf 74}, 020501
(2006).



\bibitem{LANL} Bianchi, A. Movshovich, R., Vekhter, I., Pagliuso,
P.G. \& Sarrao, J.L. Avoided antiferromagnetic order and quantum
critical point in CeCoIn$_5$. {\it Phys. Rev. Lett.} {\bf 91},
257001 (2003).


\bibitem{Sherbrooke} Paglione, J. {\it et al.} Field-induced quantum critical point in
CeCoIn$_5$. {\it Phys. Rev. Lett.} {\bf 91}, 246405 (2003).

\bibitem{Wirth} Singh, S. {\it et al.} Probing the quantum critical behavior in CeCoIn$_5$ via Hall effect
measurements. {\it Phys. Rev. Lett.} {\bf 98}, 057001 (2007).

\bibitem{Donath} Donath, J.G., Gegenwart, P., Steglich, F., Bauer, E.D. \&
Sarrao, J.L. Dimensional crossover of quantum critical behavior in
CeCoIn$_5$. arXiv:0704.0506 (2007).

\bibitem{Pham_06} Pham, L.D., Park, T., Maquilon, S., Thompson, J.D. \& Fisk, Z. Reversible tuning of the heavy-fermion ground state in
CeCoIn$_5$. {\it Phys. Rev. Lett.} {\bf 97}, 056404  (2006).


\bibitem{Gegenwart_99} Gegenwart, P. {\it et al.} Non-Fermi liquid
effects at ambient pressure in a stoichiometric heavy-fermion
compound with very low disorder: CeNi$_2$Ge$_2$. {\it Phys. Rev.
Lett.} {\bf 82}, 1293 (1999).


\bibitem{Loidl 92} Loidl, A. {\it et al.} Local-moment and itinerant
antiferromagnetism
 in the heavy-fermion system Ce(Cu$_{1-x}$Ni$_x$)$_2$Ge$_2$. {\it
 Ann. Phys.} {\bf 1}, 78-91 (1992).



%
%







\bibitem{Knafo 04} Knafo, W. {\it et al.} Anomalous scaling
behavior of the dynamical spin susceptibility of
Ce$_{0.925}$La$_{0.075}$Ru$_2$Si$_2$. {\it Phys. Rev. B} {\bf 70},
174401 (2004).

\bibitem{Montfrooij03}
Montfrooij, W. {\it et al.} Non-local quantum criticality in
Ce(Ru$_{1-x}$Fe$_x$)$_2$Ge$_2$ ($x=x_c=0.77$). {\it Phys. Rev.
Lett.} {\bf 91}, 087202 (2003).



\bibitem{Heuser 98b} Heuser, K., Scheidt, E.-W., Schreiner, T. \&
Stewart, G.R. Disappearance of hyperscaling at low temperatures in
non-Fermi liquid CeCu$_{5.2}$Ag$_{0.8}$. {\it Phys. Rev. B} {\bf
58}, 15959-15961 (1998).

\bibitem{Stockert Enderle} Stockert, O., Enderle, M. \&
L\"{o}hneysen, v.H. Magnetic fluctuations at a field-induced quantum
phase transition. {\it Phys. Rev. Lett.} {\bf 99}, 237203 (2007).







\bibitem{Belitz_rmp05} Belitz, D., Kirkpatrick, T.R. \& Vojta, T. How generic scale invariance influences quantum and classical phase
transitions. {\it Rev. Mod. Phys.} {\bf 77} 579-632 (2005).


\bibitem{Chubukov} Abanov, Ar.\ \& Chubukov, A.\ V.
Spin-fermion model near the quantum critical point: One-loop
renormalization group results. {\it Phys.\ Rev.\ Lett.} {\bf 84},
5608-5611 (2000).

\bibitem{Sullow_prl99} S\"{u}llow, S., Aronson, M.C., Rainford,
B.D. \& Haen, P. Doniach phase diagram, revisited: From ferromagnet
to Fermi liquid in pressurized CeRu$_2$Ge$_2$. {\it Phys. Rev.
Lett.} {\bf 82}, 2963-2966 (1999).


\bibitem{Fisk-ferro}
Sidorov, V.A. {\it et al.} Magnetic phase diagram of the
ferromagnetic Kondo-lattice compound CeAgSb$_2$ up to 80 kbar. {\it
Phys.~Rev.~B} {\bf 67}, 224419 (2003).

\bibitem{Lonzarich}
Lonzarich, G.\ G. in {\it Electron} (ed. Springford, M.) Ch. 6
(Cambridge Univ.\ Press, Cambridge, 1997), p. 109--147 and
references therein.


\bibitem{Lonzarich03}
Doiron-Leyraud, N. {\it et al.}  Fermi-liquid breakdown in the
paramagnetic phase of a pure metal. {\it Nature} {\bf 425}, 595-599
(2003).

\bibitem{ZrZn2} Uhlarz, M., Pfleiderer, C. \& Hayden, S.M. Quantum phase transitions in the itinerant ferromagnet
ZrZn$_2$. {\it Phys.~Rev.~Lett.} {\bf 93}, 256404 (2004).

\bibitem{FeGe}
Pedrazzini, P. {\it et al.} Metallic state in cubic FeGe beyond its
quantum phase transition. {\it Phys. Rev. Lett.} {\bf 98}, 047204
(2007).


\bibitem{Schofield-Millis} Millis, A.J., Schofield, A.J., Lonzarich, G.G. \&
Grigera, S.A. Metamagnetic quantum criticality in metals. {\it Phys.
Rev. Lett.} {\bf 88}, 217204 (2002).

\bibitem{grigera-science01}
Grigera, S.A. {\it et al.} Magnetic field-tuned quantum criticality
in the metallic ruthenate Sr{$_3$}Ru{$_2$}O{$_7$}. {\it Science}
{\bf 294}, 329-332 (2001).

\bibitem{grigera-science04}
Grigera, S.A. {\it et al.} Disorder-sensitive phase formation linked
to metamagnetic quantum criticality. {\it Science} {\bf 306},
1154-1157 (2004).


\bibitem{Flouquet02}
Flouquet, J., Haen, P., Raymond, S., Aoki, D. \& Knebel, G.
Itinerant metamagnetism of CeRu{$_2$}Si{$_2$}: Bringing out the
dead. Comparison with the new Sr{$_3$}Ru{$_2$}O{$_7$} case. {\it
Physica B} {\bf 319}, 251-261 (2002).



\bibitem{Weickert} Weickert, F. {\it et al.} Search for a quantum critical end-point in
CeRu$_2$(Si$_{1-x}$Ge$_x$)$_2$. {\it Physica B} {\bf 359}, 68-70
(2005).

\bibitem{julian06}
Daou, R., Bergemann, C. \& Julian, S.R. Continuous evolution of the
Fermi surface of CeRu{$_2$}Si{$_2$} across the metamagnetic
transition. {\it Phys.~Rev.~Lett.} {\bf 96}, 026401 (2006).


\bibitem{Kim 2003} Kim, K.H., Harrison, N., Jaime, M., Boebinger, G.S. \&
Mydosh, J.A. Magnetic-field-induced quantum critical point and
competing order parameters in URu$_2$Si$_2$. {\it Phys.~Rev.~Lett.}
{\bf 91}, 256401 (2003).

\end{thebibliography}
\end{document}